\documentclass[showpacs,preprintnumbers,eqsecnum,twocolumn,prd]{revtex4}  

\usepackage{graphicx}
\usepackage{amssymb}

\def\beq{\begin{equation}}
\def\eeq{\end{equation}}

\begin{document}

\title{Dark energy from cosmological fluids obeying a Shan-Chen non-ideal equation of state}

\author{Donato Bini}
  \affiliation{
Istituto per le Applicazioni del Calcolo ``M. Picone,'' CNR, I-00185 Rome, Italy\\
ICRA, ``Sapienza" University of Rome, I-00185 Rome, Italy\\
INFN, Sezione di Firenze, Polo Scientifico, Via Sansone 1, 50019 Sesto Fiorentino, Florence, Italy
}

\author{Andrea Geralico}
  \affiliation{Physics Department and ICRA, ``Sapienza" University of Rome, I-00185 Rome, Italy}

\author{Daniele Gregoris}
  \affiliation{Department of Physics, Stockholm University, 106 91 Stockholm, Sweden\\
ICRA, ``Sapienza" University of Rome, I-00185 Rome, Italy\\
Max-Planck-Institut f\"ur Gravitationsphysik (Albert-Einstein-Institut) Am M\"uhlenberg 1, DE-14476 Potsdam, Germany
}

\author{Sauro Succi}
  \affiliation{Istituto per le Applicazioni del Calcolo ``M. Picone,'' CNR, I-00185 Rome, Italy\\
INFN, Sezione di Firenze, Polo Scientifico, Via Sansone 1, 50019 Sesto Fiorentino, Florence, Italy}

\begin{abstract}
We consider a Friedmann-Robertson-Walker universe with a fluid source obeying a 
non-ideal equation of state with \lq\lq asymptotic freedom," namely ideal gas behavior (pressure changes
directly proportional to density changes) both at low and high density regimes, following a fluid dynamical model due to Shan and Chen.
It is shown that, starting from an ordinary energy density component, such fluids naturally evolve towards a 
universe with a substantial \lq\lq dark energy" component at the present time, with no need of invoking
any cosmological constant. 
Moreover, we introduce a quantitative indicator of darkness abundance, which provides a consistent
picture of the actual matter-energy content of the universe.  
\end{abstract}


\maketitle

\section{Introduction}

Current improvements in cosmological measurements strongly favor the standard model of the universe being spatially flat, homogeneous and isotropic on large scales and dominated by dark energy consistently with the effect of a cosmological constant and cold dark matter.
Such a concordance model is referred to as $\Lambda$-Cold Dark Matter ($\Lambda$CDM) model in the literature and depends on six cosmological parameters: the density of dark matter, the density of baryons, the expansion rate of the universe, the amplitude of the primordial fluctuations, their scale dependence, and the optical depth of the universe.
These parameters are enough to successfully describe all current cosmological data sets, including the measurements of temperature and polarization anisotropy in the cosmic microwave background (CMB) (see, e.g., Ref. \cite{hinshaw} and references therein).
Therefore, according to the $\Lambda$CDM model the universe is well described by a Friedmann-Robertson-Walker (FRW) metric, whose gravity source is a mixture of non-interacting perfect fluids including a cosmological constant.

Observations of distant type Ia supernovae (SNe Ia) first pointed to the so-called {\it dark energy} as 
a major actor in driving the accelerated expansion of the universe \cite{riess98,perlmutter99}. 
Combined observations of large scale structure and the cosmic microwave background radiation then provided indirect evidence of a dark energy component with negative pressure, which gives the dominant contribution to the whole mass-energy content of the universe (see, e.g., Refs. \cite{ratra,spergel,padman}). 
At present, all existing observational data are in agreement with the simplest picture of
dark energy as a cosmological constant effect, i.e. the $\Lambda$CDM model.
Nevertheless, no theoretical model determining the nature of dark energy is 
available as yet, leaving its existence still unexplained.
Other possibilities of a (slightly) variable dark energy have also been considered in recent years.
These models include, for instance, a decaying scalar field (quintessence) minimally coupled to gravity, similar 
to the one assumed by inflation \cite{ratra}, scalar field models with nonstandard kinetic terms ($k-$essence) \cite{apicon}, the Chaplygin gas \cite{bento}, braneworld models and cosmological models from scalar-tensor theories of gravity (see, e.g., Refs. \cite{sahni,lapuente} and references therein).
The possibility that the acceleration of the universe could be driven by the bulk viscosity of scalar theories has also been explored \cite{padmachitre}. 
Relaxation processes associated with viscous fluid have been shown to reduce the effective pressure, which could become negative for a sufficiently large bulk viscosity, so mimicking a dark energy behavior \cite{gagnon}. 
The present paper falls in the line of cosmological models with modified equation of state \cite{brevik}.
The main idea is  to postulate that the cosmological fluid obeys a non-ideal equation of state 
with ``asymptotic freedom,'' namely  ideal gas behavior (pressure and density changes in linear
proportion to each other) at both low and high density regimes, with
a liquid-gas coexistence loop in between. 
Such non-ideal equation state supports a phase transition, which models the growth
of the dark matter-energy component of the universe, as a natural consequence of the fluid evolution equations. 

The idea of an asymptotic-free, non-ideal equation of state was first proposed 
by Shan and Chen (SC) in the context of lattice kinetic theory, with the primary intent of 
producing a liquid-vapor coexistence curve with {\it purely attractive} interactions \cite{shan-chen} (see Appendix A).
Its distinctive feature is to replace hard-core repulsive interactions, as needed to 
tame unstable density build-up, with a purely attractive force, with the peculiar
property of becoming vanishingly small above a given density threshold, i.e. 
a form of effective ``asymptotic freedom'' \cite{ASYF}.
The SC motivation was purely numerical, namely do away with the very small time-steps
imposed by the hard-core repulsion in the numerical integration of the lattice kinetic equations.
Indeed, in the last two decades, the SC method has met with major success for the numerical 
simulation of a broad variety of complex flows with phase-transitions \cite{LB1,LB2}.  

In this work, we maintain that the peculiar properties of the SC equation of state
may offer fresh new insights into cosmological fluid dynamics, and most notably for
the development of a new class of cosmological models with scalar gravity. 
In particular, given that the SC approach has proven very successful in 
dispensing with hard-core repulsion in ordinary fluids, it might be envisaged 
that, in the cosmological context, it would permit to do away with the repulsive 
action of the cosmological constant.
 
As we shall see, this is just the case: a cosmological FRW fluid obeying the SC equation of state naturally evolves towards a present-day universe with a suitable dark-energy component, with no need of invoking any cosmological constant.

\section{Basic equations of the model}

The Friedmann-Robertson-Walker metric written in comoving coordinates is given by \cite{exact} 
\beq
ds^2=-dt^2+a^2\left[dr^2+\Sigma_k^2(d\theta^2+\sin^2\theta d\phi^2) \right]\,, 
\eeq
where $a=a(t)$ is the scale factor and $\Sigma_k=\Sigma_k(r)=[\sin r, r, \sinh r]$ corresponding to closed, flat and open universes, respectively.
The matter-energy content of the universe is assumed to be a perfect fluid at rest with respect to the coordinates (i.e., with $u=\partial_t$ as the fluid 4-velocity, $u_\alpha u^\alpha=-1$) satisfying a Shan-Chen-like equation of state, i.e.,
\begin{eqnarray}
\label{pscdef}
p_{\rm (sc)}&=&w_{\rm (in)}\rho_{\rm (crit),0}  \left[\frac{\rho}{\rho_{\rm (crit),0} }+\frac{g}{2} \psi^2\right]\,,\nonumber\\
\psi&=& 1-e^{-\alpha \frac{\rho}{\rho_{\rm (crit),0}}}\,,
\end{eqnarray}
where $\rho_{\rm (crit),0}={3H_0^2}/{8\pi}$ is the present value of the critical density ($H_0$ denoting the Hubble constant) and the dimensionless quantities $w_{\rm (in)}$, $g \le 0$ and $\alpha\ge0$ can be regarded as free parameters of the model. 
A short review of the original Shan-Chen model is presented in Appendix A.
Notice that in principle one should have written $
\psi \propto 1-e^{-\frac{\rho}{\rho_*}}$, 
$\rho_*$  being the typical density  above which  $\psi$ undergoes a \lq\lq saturation effect," $\psi\approx 1$. 
Equivalently, here  we have denoted $\rho_*=\rho_{\rm (crit),0}/\alpha$, and expressed the saturation scale 
in terms of the free parameter $\alpha$.

The quantity $\psi$ can be interpreted as the density of a chameleon scalar field \cite{khoury}, reducing to ordinary matter, i.e., $\psi \to \rho$, in the low density limit $\rho \ll \rho_*$ and asymptotically goes to a uniform constant in the opposite limit.
This scalar field carries a purely attractive interaction and consequently it contributes a negative pressure to the equation of state.
Since the associated force vanishes in the limit $\rho \gg \rho_*$, this regime corresponds to an effective form of ``asymptotic freedom,'' occurring at cosmological rather than subnuclear scales.
Similarly to the case of lattice kinetic theory, in which the stabilizing effect 
of hard-core repulsion is replaced by an asymptotic-free attraction, the repulsive effect of the cosmological 
constant is here replaced by a scalar field with asymptotic-free attraction. 
At present, the existence of such an extra scalar field cannot be taken for more than a speculation, but we will show below that such a speculation permits to interpret actual cosmological data in a very natural and elegant way, with no need of invoking any cosmological constant.

The associated stress-energy tensor is given by 
\beq
\label{stress-energy_sc}
T^{\rm (sc)}_{\alpha\beta}= (\rho+p_{\rm (sc)})u_\alpha u_\beta+p_{\rm (sc)}g_{\alpha\beta}\,, 
\eeq
where a self-pressure-induced contribution to the energy density  (SC pressure hereafter, $p_{\rm (sc)}$, related to $\rho$ by Eq. (\ref{pscdef})) arises as a typical feature of the model. 
The evolution of the energy density $\rho=\rho(t)$ is obtained from Einstein's field equations $G_{\mu\nu}+\Lambda g_{\mu\nu}=8\pi T_{\mu\nu}$, which in this case can be reduced to the energy conservation equation 
\beq
\label{energy}
\dot\rho=-3\frac{\dot a}{a}(\rho +p_{\rm (sc)})\,,
\eeq
and the Friedmann equation
\beq
\label{friedeq}
\dot a^2  = -k +\frac83 \pi \rho  a^2+\frac{\Lambda}{3}a^2\,,
\eeq
where $k=[1,0,-1]$ for the case of closed, flat and open universes, respectively.
Dot and prime denote derivative with respect to time and $r$, respectively.
The Friedmann equation (\ref{friedeq}) can be equivalently rewritten in terms of Hubble parameter $H=\dot a/a$ and critical density $\rho_{\rm (crit)}={3H^2}/{8\pi}$ as
\beq
\label{friedeq3}
H^2\equiv\frac{\dot a^2}{a^2}  = -\frac{k}{a^2} + H^2\frac{\rho}{\rho_{\rm (crit)}}+\frac{\Lambda}{3}\,.
\eeq
Introducing then the SC density parameter $\Omega_{\rm (sc)}$, the curvature parameter $\Omega_k$ and the vacuum energy parameter $\Omega_\Lambda$ defined by
\beq
\label{defOmegas}
\Omega_{\rm (sc)}=\frac{\rho}{\rho_{\rm (crit)}}\,, \qquad
\Omega_k=-\frac{k}{H^2a^2}\,,\qquad 
\Omega_\Lambda=\frac{\Lambda}{3H^2} \,,
\eeq
Eq. (\ref{friedeq3}) takes the simple form
\beq
\label{friedeq4}
\Omega_{\rm (sc)}+\Omega_k+\Omega_\Lambda=1\,.
\eeq
The corresponding present-day values (at $t=t_0$) will be denoted by a subscript ``0." 
It is also useful to introduce the deceleration parameter $q=-\ddot a/(aH^2)$ with the associated acceleration equation
\beq
\label{accel}
\frac{\ddot a}{a}=-\frac{4\pi}{3}(\rho +3p_{\rm (sc)})+\frac{\Lambda}{3}\,,
\eeq
which describes the acceleration of the scale factor (it is obtained from both Friedmann and fluid equations), so that
\beq
\label{qdef}
q=\frac{\Omega_{\rm (sc)}}{2}+\frac32\frac{p_{\rm (sc)}}{\rho_{\rm (crit)}}-\Omega_\Lambda\,.
\eeq

\subsection{General features}

In order to investigate the general features of Shan-Chen cosmologies it is convenient to cast the model equations in a form which is suited to numerical integration by introducing the following set of dimensionless variables 
\beq
\xi=\frac{\rho}{\rho_{\rm (crit),0}}\,, \qquad
x=\frac{a}{a_0}\,, \qquad 
\tau=H_0t\,.
\eeq
The dimensionless density $\xi$ is related to the SC density parameter $\Omega_{\rm (sc)}$ introduced in Eq. (\ref{defOmegas}) by 
\beq
\label{xi_vs_omega}
\xi=\Omega_{\rm (sc)}\frac{H^2}{H_0^2}\,.
\eeq 
The equation of state (\ref{pscdef}) thus assumes the (rescaled) simplified form
\begin{eqnarray}
p_{\rm (sc)}&=&w_{\rm (in)}\rho_{\rm (crit),0}  \left[\xi +\frac12 g \left(1-e^{-\alpha \xi }\right)^2\right]\nonumber\\
&\equiv&w_{\rm (in)}\rho_{\rm (crit),0}  {\mathcal P}_{(\alpha,g)}(\xi)\,,
\end{eqnarray}
so that the SC pressure has the same sound speed 
\beq
\label{csdef}
c_s^2 \equiv \frac{\partial p_{\rm (sc)}} {\partial \rho}
=w_{\rm (in)}\left[1+g\alpha\left(1-e^{-\alpha \xi }\right)e^{-\alpha \xi }\right]
\eeq
both in the low and high density limits, where $c_s^2\to w_{\rm (in)}$ for fixed values of $\alpha$.
In fact, for $\xi\ll1$ the function ${\mathcal P}_{(\alpha,g)}(\xi)\sim\xi$, and for $\xi\gg1$ it goes to ${\mathcal P}_{(\alpha,g)}(\xi)\sim \xi+g/2$, which looks exactly like the bag-model equation of state of hadronic matter \cite{BAG}.
Here, $g<0$ plays the role of the bag constant, i.e., the difference between the energy density of the true vacuum versus the perturbative one. 
We refer to the high-density branch as to ``ideal gas" behavior, in the sense that pressure and density changes are directly proportional to each other. 
On the other hand, varying the parameter $\alpha$ in the allowed range the SC pressure behaves again as $p_{\rm (sc)}\sim w_{\rm (in)}\rho$ for $\alpha \to 0$, whereas for $\alpha \to \infty$   
\beq
p_{\rm (sc)}\sim w_{\rm (in)}\rho +\Delta p\,,\qquad
\Delta p =\frac{g}{2} w_{\rm (in)} \rho_{\rm (crit),0}\,.
\eeq

The energy conservation equation (\ref{energy}) and the Friedmann equation (\ref{friedeq}) can then be written as
\begin{eqnarray}
\label{eqsvstau1}
\frac{d\xi}{d\tau }&=&
-\frac{3}{x}\frac{dx}{d\tau}\left[(1+w_{\rm (in)})\xi\right.\nonumber\\
&&\left. + \frac{w_{\rm (in)}}2g\left(1-e^{-\alpha \xi}  \right)^2  \right]\,,\\
\label{eqsvstau2}
\frac{dx}{d\tau}&=&\pm\sqrt{\Omega_{k,0}+x^2(\xi+\Omega_{\Lambda,0})}\,,
\end{eqnarray}
which can be numerically integrated with initial conditions $\xi(\tau_0)=\Omega_{{\rm (sc)},0}$ and $x(\tau_0)=1$.
The $\pm$ sign in front of the rhs of the second equation corresponds to increasing ($+$)/decreasing ($-$) behavior of the scale factor.
The value of $\tau_0$ must be chosen in such a way that the numerical integration gives the correct behavior of the solution approaching the initial singularity, i.e., $x\to0$ for $\tau\to0$.

Note that Eq. (\ref{eqsvstau1}) can also be formally integrated to give $x=x(\xi)$ as
$$
x=e^{-{\mathcal L}(\xi)}\,,\qquad
{\mathcal L}(\xi)=\frac13\int_{\Omega_{{\rm (sc)},0}}^{\xi}\frac{d\eta}{\eta+w_{\rm (in)}{\mathcal P}_{(\alpha,g)}(\eta)}\,.
$$
The evolution $x=x(\tau)$ then follows from Eq. (\ref{eqsvstau2}).

\begin{figure}
\begin{center}
\includegraphics[scale=0.4]{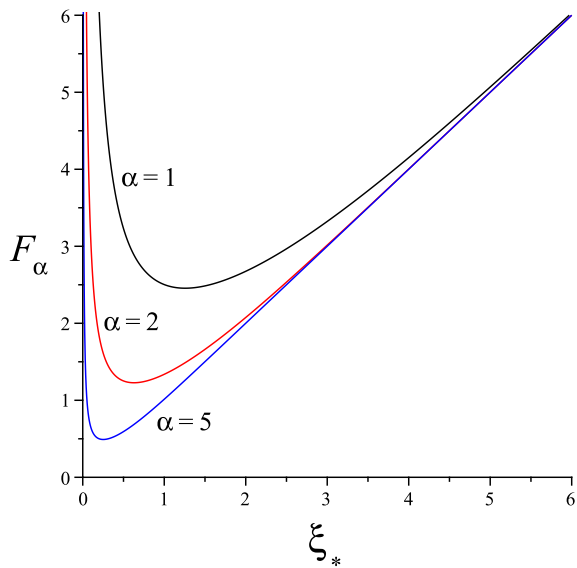}
\end{center}
\caption{
The function $F_\alpha(\xi_*)$ determining the equilibrium solutions in the case of a flat universe (see Eq. (\ref{eq_F})) is plotted for different values of $\alpha=[1,2,5]$. 
The existence of two solutions is evident by drawing horizontal lines. 
$F_\alpha$ has a minimum at 
$\xi_*^{\rm min}=-\frac{1}{2\alpha}\left[2W_{-1}\left(-\frac{1}{2\sqrt{e}} \right)+1\right]\approx1.256/\alpha$, with value $F_\alpha(\xi_*^{\rm min})\approx2.455/\alpha$. 
Here $W_k(z)$ denotes the $k-$branch of the Lambert $W$ function satisfying the equation $W(z)e^{W(z)}= z$, where $k$ is any non-zero integer \cite{corless}.
If the variable $z$ is real, then there are two possible real values of $W(z)$ in the interval $-1/e\leq z <0$.
The branch satisfying $W(z)\geq -1$ is denoted by $W_0(z)$ and is referred to as the principal branch of the $W$ function, whereas the branch satisfying $W(z)\leq-1 $ is denoted by $W_{-1}(z)$.
The existence of two roots for Eq. (\ref{eq_F}) is thus guaranteed if $2.455/\alpha\lessapprox{w_{\rm (in)}|g|}/{2(1+w_{\rm (in)})}$. 
}
\label{fig:1}
\end{figure}

The system (\ref{eqsvstau1})--(\ref{eqsvstau2}) admits as equilibrium solutions $(\xi_*,x_*)$ the pair of constants satisfying the conditions ${d\xi}/{d\tau }=0$ and ${dx}/{d\tau}=0$.
Such solutions do exist in the flat case ($\Omega_{k,0}=0$) and for positively curved (i.e., closed) universes only.
In the latter case the equilibrium is characterized by arbitrary values of $\xi_*$ and 
\beq
x_*=\sqrt{\frac{|\Omega_{k,0}|}{\xi_*+\Omega_{\Lambda,0}}}\,.
\eeq
For a flat universe, Eqs. (\ref{eqsvstau1})--(\ref{eqsvstau2}) lead to the single equation
\beq
\frac{d\xi}{d\tau }=
-3\sqrt{\xi+\Omega_{\Lambda,0}}\left[(1+w_{\rm (in)})\xi + \frac{w_{\rm (in)}}2 g \left(1-e^{-\alpha \xi}  \right)^2  \right]\,.
\eeq
In this case, besides $\xi_*=-\Omega_{\Lambda,0}$, there exist in general two different equilibrium solutions such that
\beq
\label{eq_F}
F_\alpha(\xi_*)\equiv \frac{\xi_*}{\left(1-e^{-\alpha \xi_*}  \right)^2} = \frac{w_{\rm (in)}|g|}{2(1+w_{\rm (in)})}  \,,
\eeq
for fixed values of $\alpha$, as shown in Fig. \ref{fig:1}.


\subsection{Dark energy without vacuum energy}

The most likely cosmology describing the universe ($\Lambda$CDM model) has a nearly spatially flat geometry ($\Omega_{k,0}=-0.010\pm0.005$) with a matter density (dark matter plus baryons) of $\Omega_{m,0}=0.266\pm0.029$ and a cosmological constant responsible of a dark energy density of $\Omega_{\Lambda,0}=0.734\pm0.029$ \cite{komatsu}.
At early times the universe was radiation-dominated, but the present contribution of radiation is negligibly small.
The dominant contribution to the mass-energy budget of the universe today is due to dark energy, obeying an equation of state $p_{\rm de}\simeq-\rho_{\rm de}$, i.e., with $w_{\rm de}\simeq-1$.
The cosmological constant thus acts as an effective negative pressure, allowing the total energy density 
of the universe to remain constant even though the universe expands.
We show below that a simple SC model does not need any cosmological constant to account for the presence of dark energy today.

\begin{figure*}
\begin{center}
$\begin{array}{cc}
\includegraphics[scale=0.25]{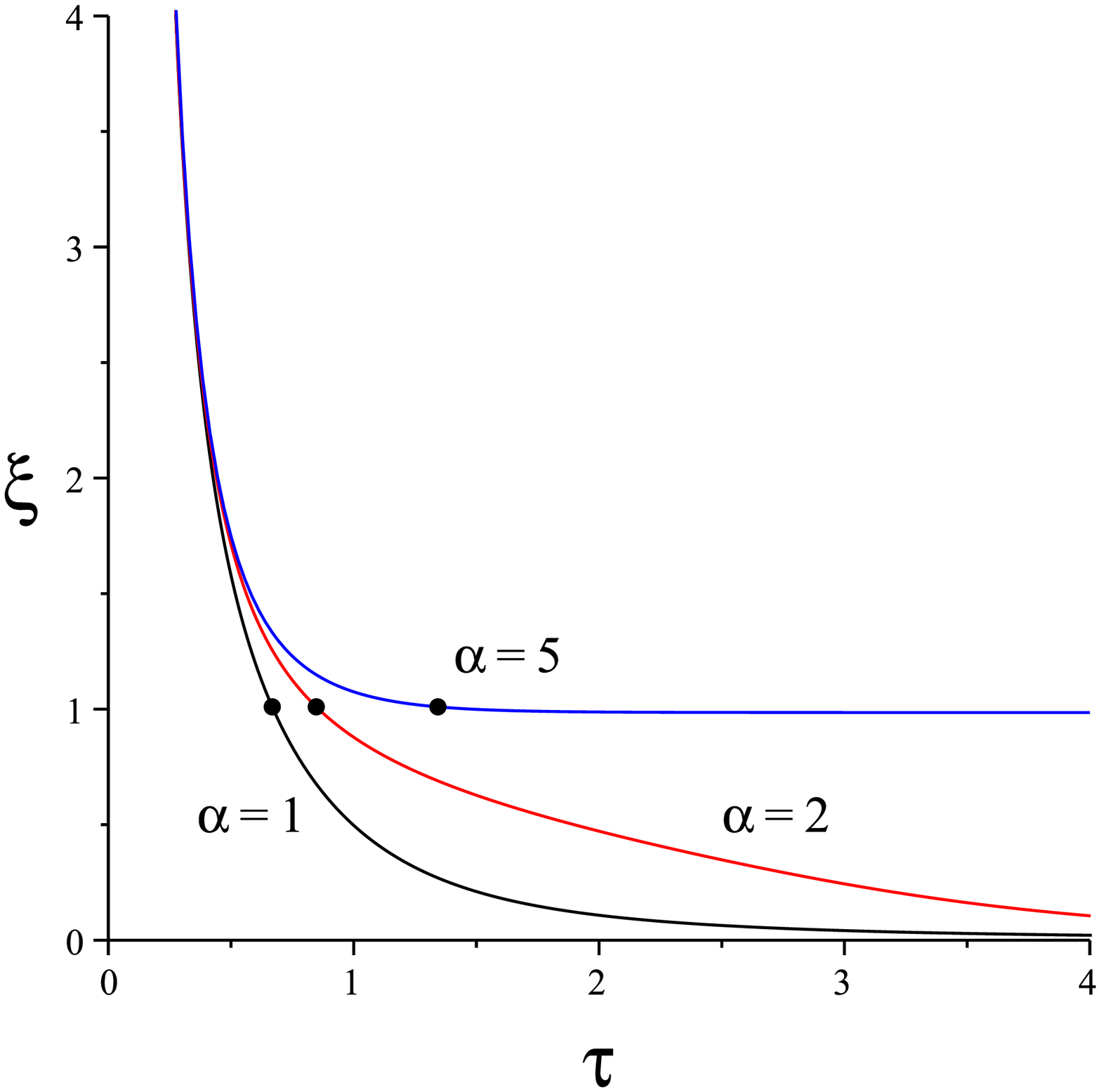}&\qquad
\includegraphics[scale=0.25]{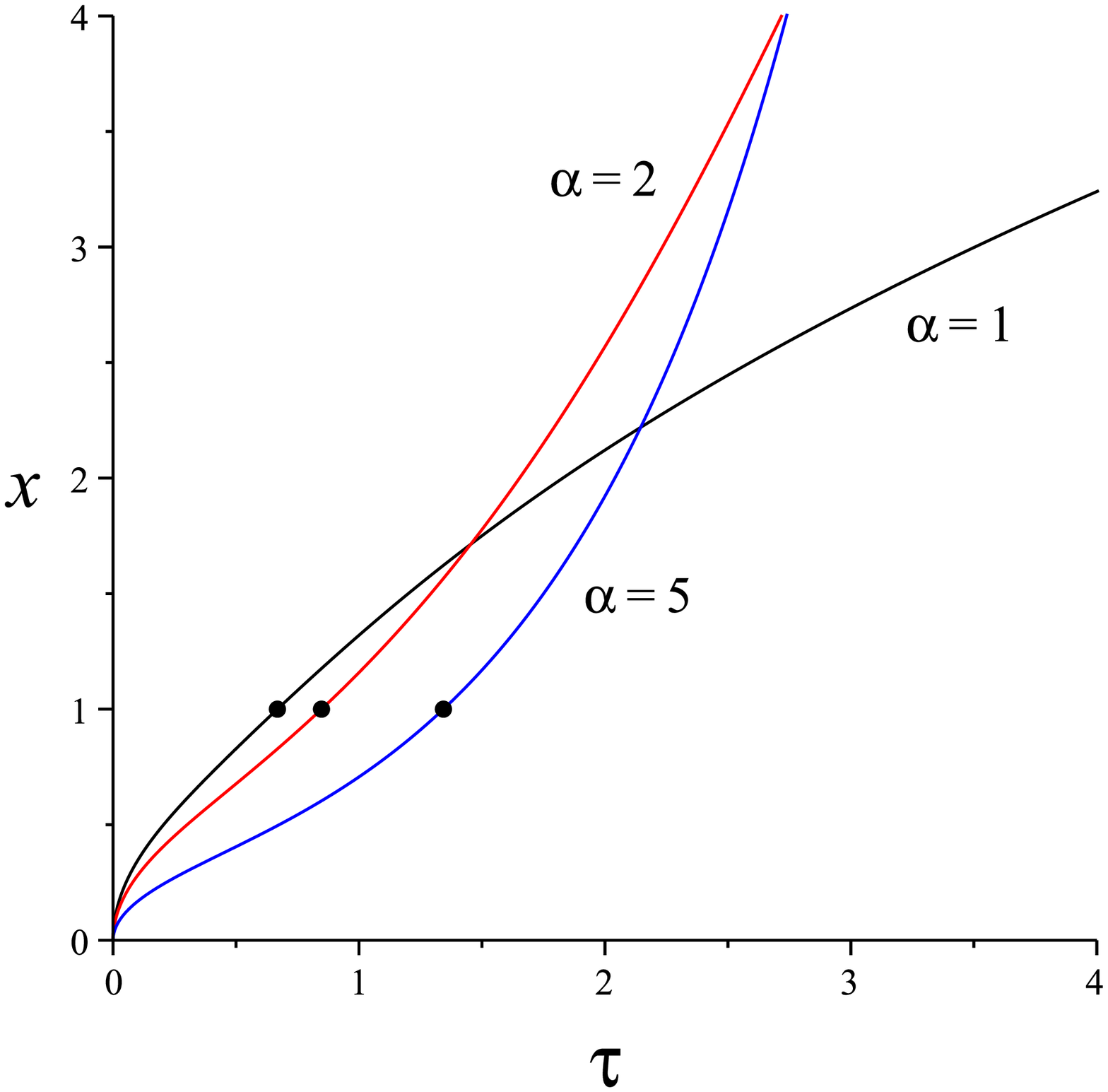}\\[.2cm]
\mbox{(a)} &\qquad \mbox{(b)}\cr
\includegraphics[scale=0.28]{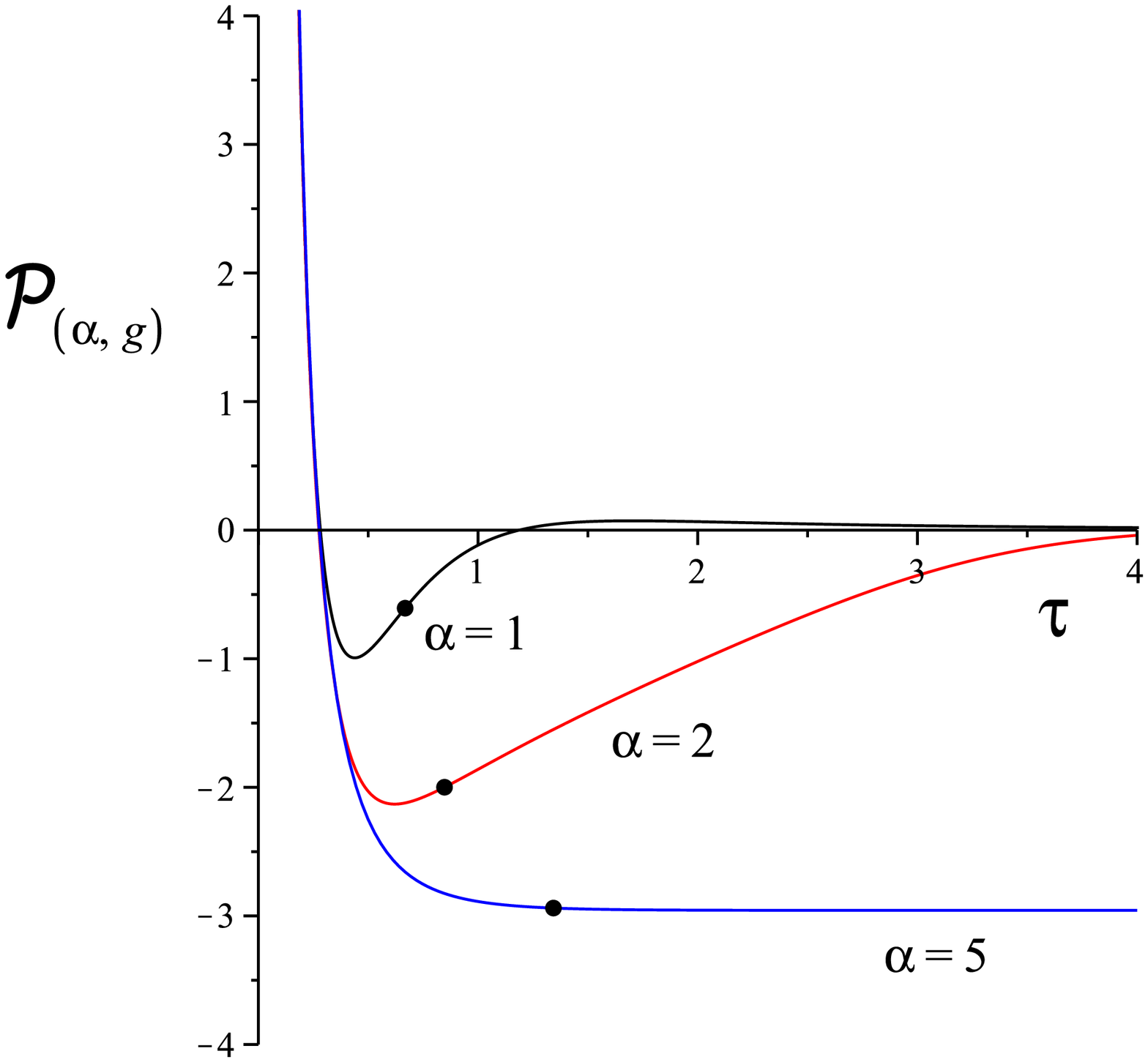}&\qquad
\includegraphics[scale=0.26]{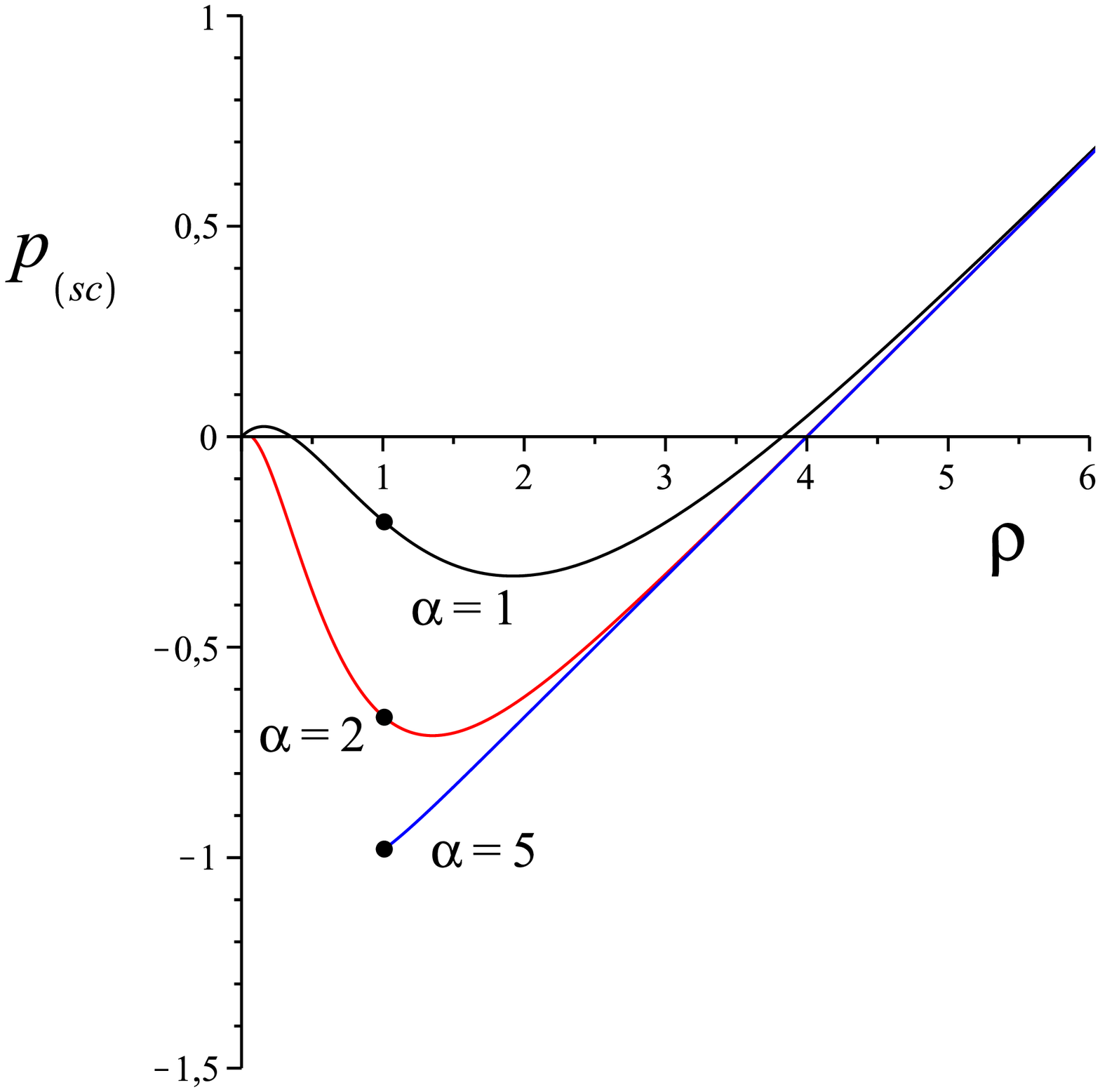}\\[.2cm]
\mbox{(c)} &\qquad \mbox{(d)}\cr
\end{array}
$\\
\end{center}
\caption{
The solutions for $\xi$ [panel (a)],  $x$  [panel (b)], and ${\mathcal P}_{(\alpha, g)}$ [panel (c)] are shown as functions of $\tau$ for the choice of parameters $w_{\rm (in)}=1/3$, $g=-8$, $\Omega_{k,0}=-0.01$, $\Omega_{\Lambda,0}=0$ and different values of $\alpha=[1,2,5]$, with initial conditions $\xi(\tau_0)=1.01$ and $x(\tau_0)=1$.
The corresponding values of $\tau_0$ are $\tau_0\approx[0.668,0.847,1.343]$, respectively. 
Panel (d) shows instead the behavior of $p_{\rm (sc)}$ as a function of $\rho$ in units of $\rho_{\rm(crit),0}$.
Note that for $\alpha=5$ the integration stops approaching the equilibrium solution $\xi\sim1$ and ${\mathcal P}_{(\alpha, g)}\sim-3$, implying $w_{\rm eff}\equiv p_{\rm (sc)}/\rho\sim-1$.
Here (and below) a dot on each curve marks the corresponding present-day value. 
}
\label{fig:2}
\end{figure*}

Consider for instance the case of an initially radiation-dominated universe, i.e., with $w_{\rm (in)}=1/3$.
The coupled set of equations (\ref{eqsvstau1})--(\ref{eqsvstau2}) is numerically integrated for a fixed value of the parameter $g$ and different values of $\alpha$, by assuming $\Omega_{\Lambda,0}=0$ (i.e., $\Lambda=0$) and $\Omega_{k,0}=-0.01$, so that $\Omega_{{\rm (sc)},0}=1-\Omega_{k,0}=1.01$.
The evolution with time of dimensionless density $\xi$, scale factor $x$ and SC pressure is shown in Fig. \ref{fig:2} for different values of $\alpha$.
We see that pressure changes its sign at a certain time in the past and remains negative on a large time interval, including 
the present epoch, so that the equation of state governing the evolution of the present universe is typical of dark energy. 
The case $\alpha=5$ exhibits a saturation effect, with the equilibrium solution $\xi\sim1$ and ${\mathcal P}_{(\alpha, g)}\sim-3$ being eventually reached during the evolution, leading to an effective $w_{\rm eff}\equiv p_{\rm (sc)}/\rho\sim-1$.


\subsection{Including a matter component}

\begin{figure*}
\begin{center}
$\begin{array}{cc}
\includegraphics[scale=0.25]{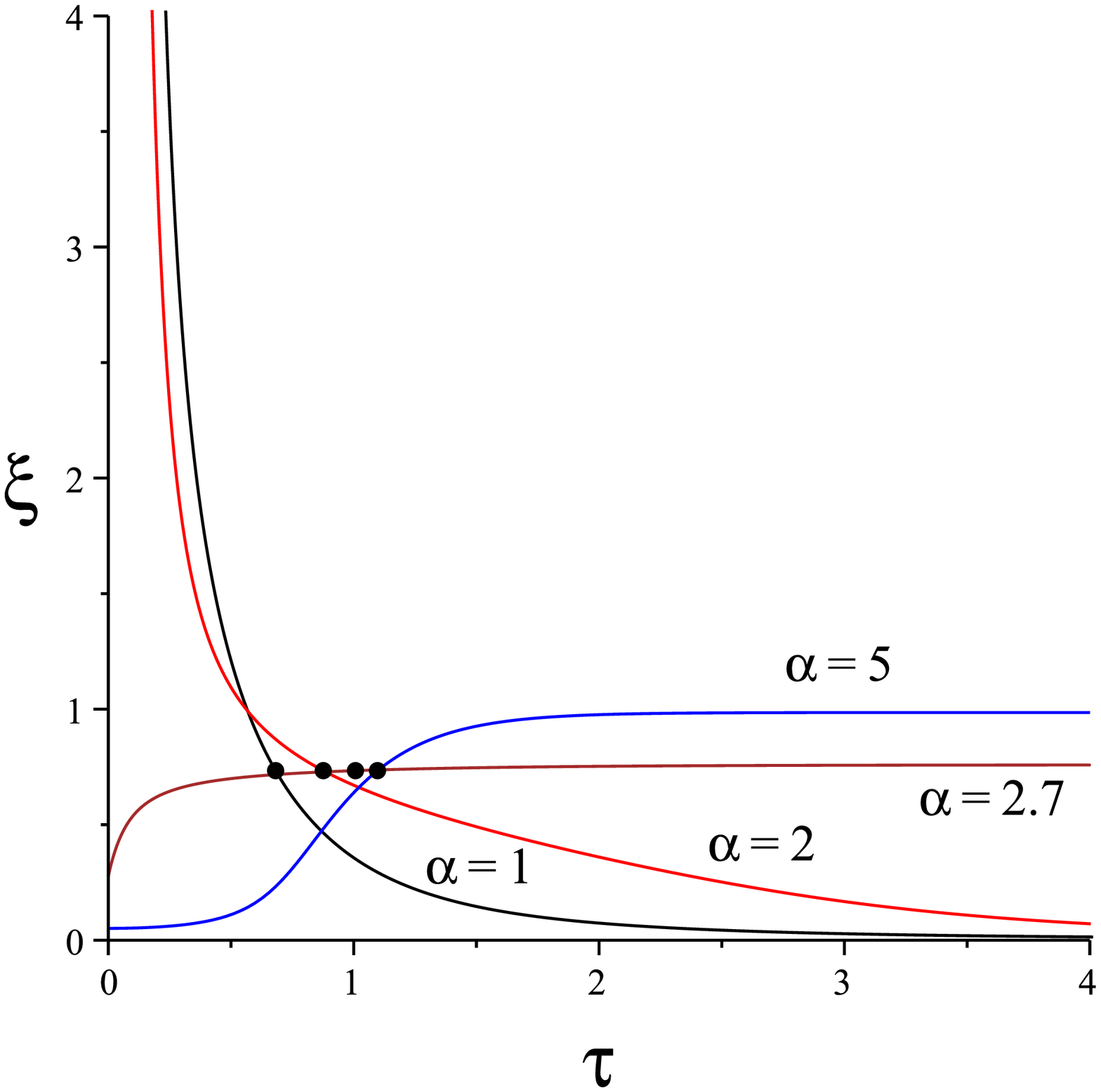}&\qquad
\includegraphics[scale=0.25]{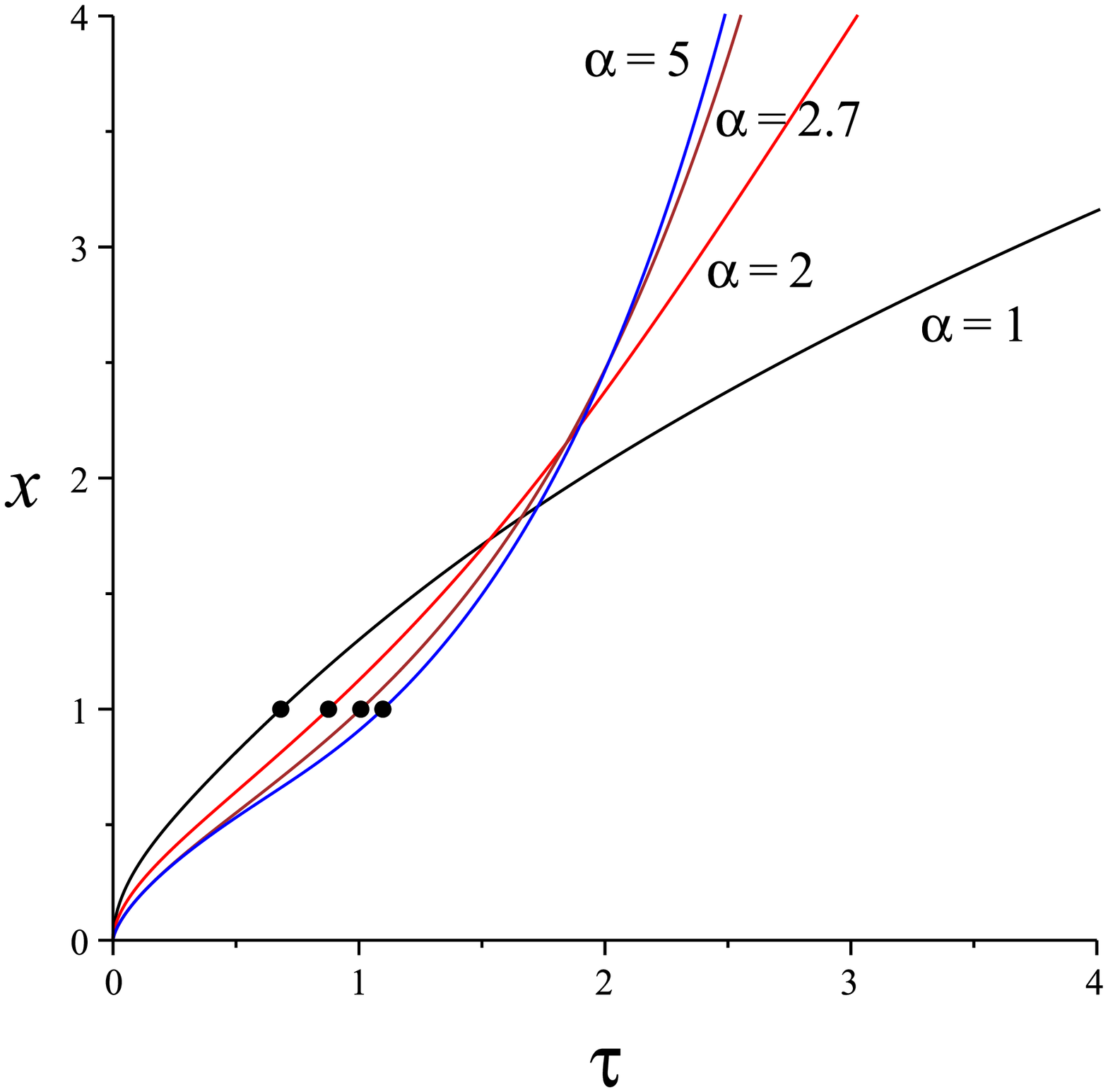}\\[.2cm]
\mbox{(a)} &\qquad \mbox{(b)}\cr
\includegraphics[scale=0.28]{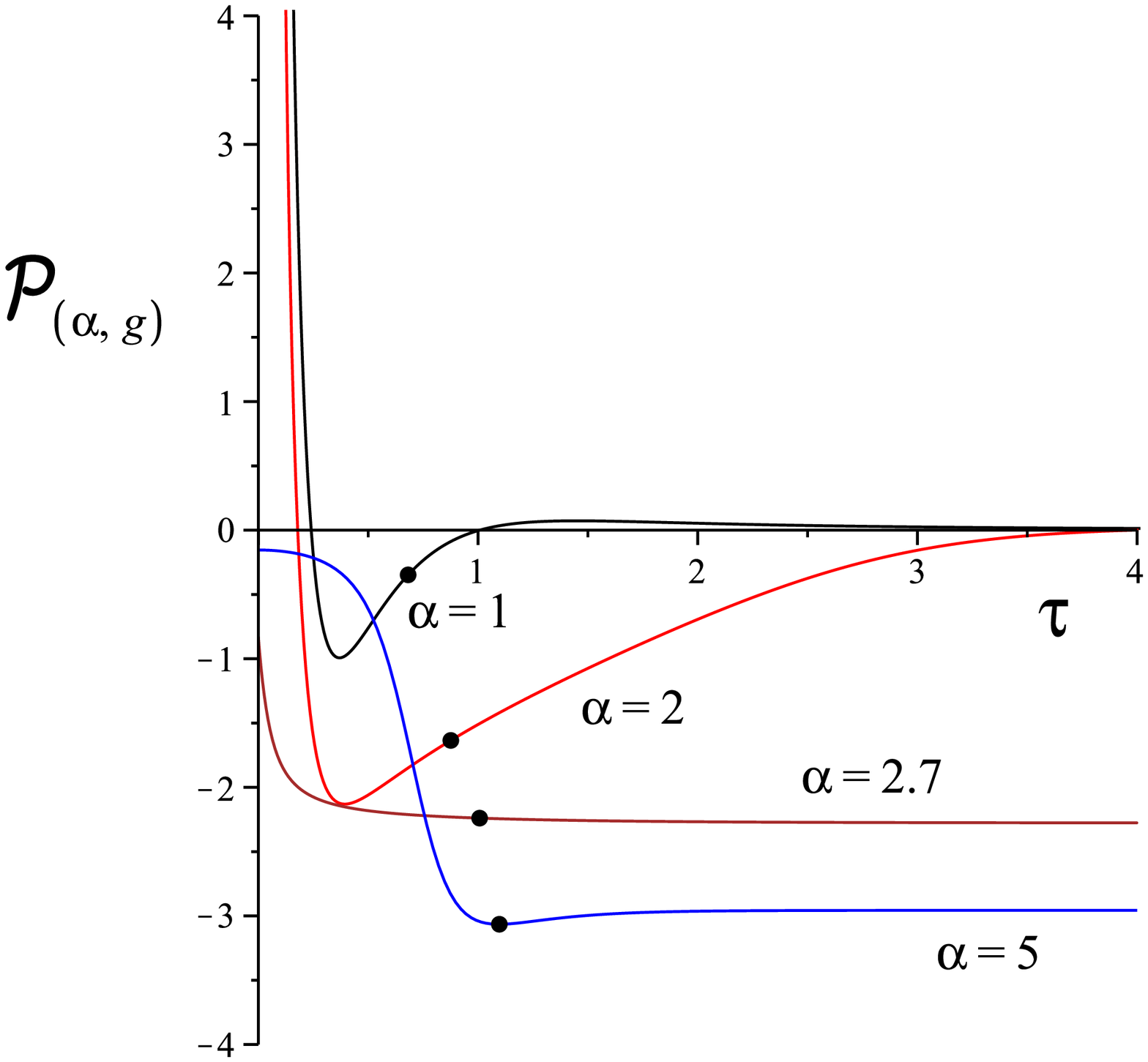}&\qquad
\includegraphics[scale=0.26]{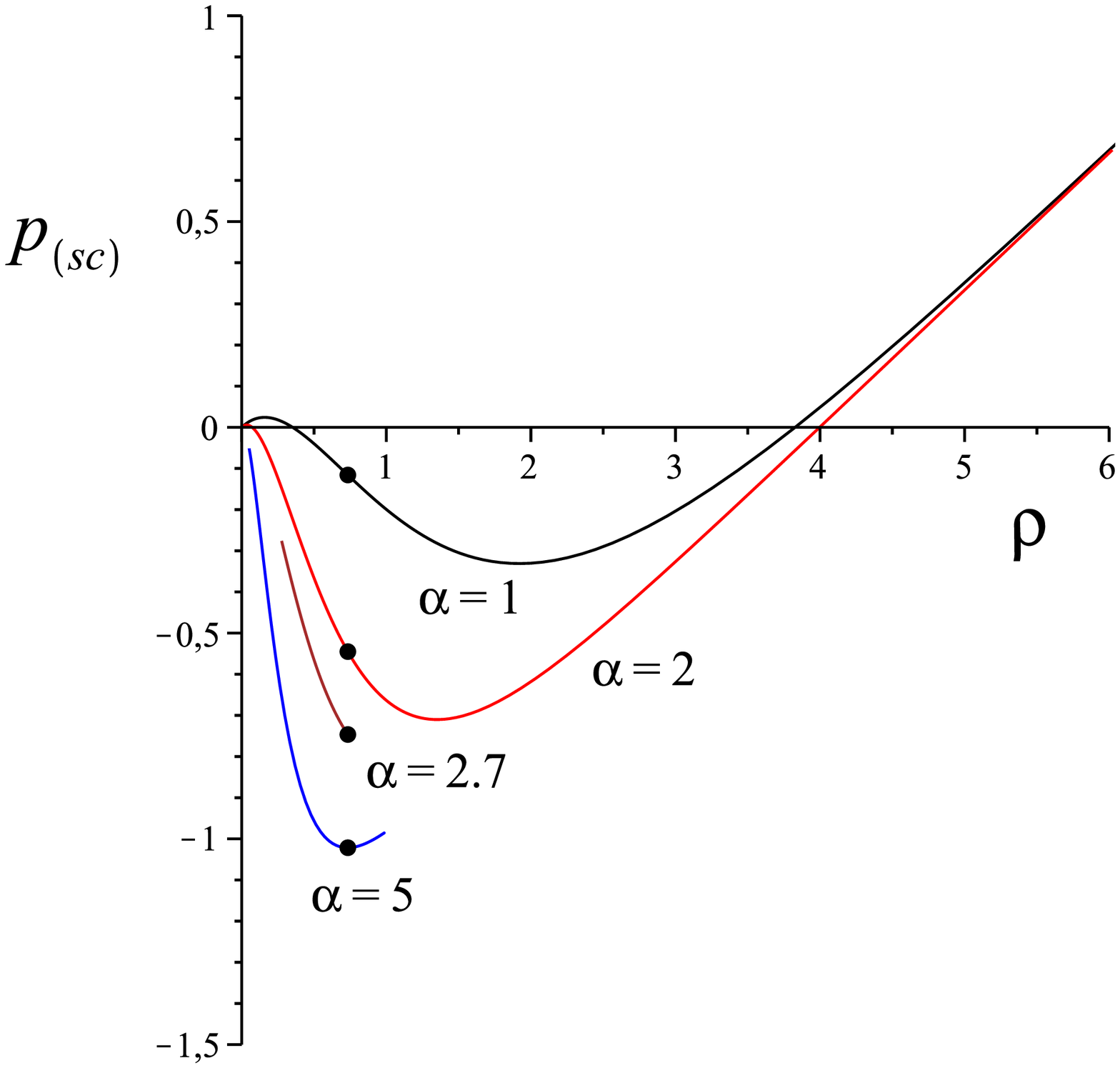}\\[.2cm]
\mbox{(c)} &\qquad \mbox{(d)}\cr
\end{array}
$\\
\end{center}
\caption{The evolution of dimensionless SC density, scale factor and SC pressure are shown in panels (a) to (c), respectively, for the choice of parameters $w_{\rm (in)}=1/3$, $g=-8$, $\Omega_{\Lambda,0}=0$, $\Omega_{m,0}=0.266$, $\Omega_{k,0}=-0.01$ and different values of $\alpha=[1,2,2.7,5]$, with initial conditions $\xi(\tau_0)=0.734$ and $x(\tau_0)=1$.
The corresponding values of $\tau_0$ are $\tau_0\approx[0.682,0.876,1.007,1.097]$, respectively. 
Panel (d) shows instead the behavior of $p_{\rm (sc)}$ as a function of $\rho$ in units of $\rho_{\rm(crit),0}$.
Note that for $\alpha=2.7$ and $\alpha=5$ the integration stops approaching the equilibrium solutions.
}
\label{fig:3}
\end{figure*}

In order to account for the presence of matter density today one has to add to the SC fluid the contribution due to pressureless matter, i.e.,
\beq 
\label{matterden}
\rho_m = \rho_{m,0} \left(\frac{a_0}{a}\right)^3\,,
\eeq
with associated density parameter
\beq 
\Omega_m = \frac{\Omega_{m,0}}{x^3}\frac{H_0^2}{H^2}\,,
\eeq
so that the Friedmann equation (\ref{friedeq4}) becomes
\beq
\label{friedeq5}
\Omega_{\rm (sc)}+\Omega_m+\Omega_k+\Omega_\Lambda=1\,,
\eeq
with $\Omega_{\rm (sc)}=\xi({H_0^2}/{H^2})$, as from Eq. (\ref{xi_vs_omega}).
The evolution equation (\ref{eqsvstau2}) for the dimensionless scale factor is thus replaced by
\beq
\label{eqsvstau2new}
\frac{dx}{d\tau}=\pm\sqrt{\Omega_{k,0}+x^2(\xi+\Omega_{\Lambda,0})+\frac{\Omega_{m,0}}{x}}\,.
\eeq
The results of the numerical integration of the system (\ref{eqsvstau1}) and (\ref{eqsvstau2new}) are shown in Fig. \ref{fig:3} for the choice of density parameters $\Omega_{\Lambda,0}=0$, $\Omega_{m,0}=0.266$ and $\Omega_{k,0}=-0.01$ and different values of $\alpha$.
The evolution of SC density exhibits a twofold behavior as a function of $\alpha$.
For $\alpha\gtrsim2$ it indefinitely grows as the initial singularity is approached, while it vanishes at late times.
As $\alpha$ increases, there exists a critical value of $\alpha$ above which the density reaches an equilibrium solution by integrating both backward and forward in time, i.e., it evolves between two equilibrium states.

\begin{figure*}
\begin{center}
$\begin{array}{cc}
\includegraphics[scale=0.3]{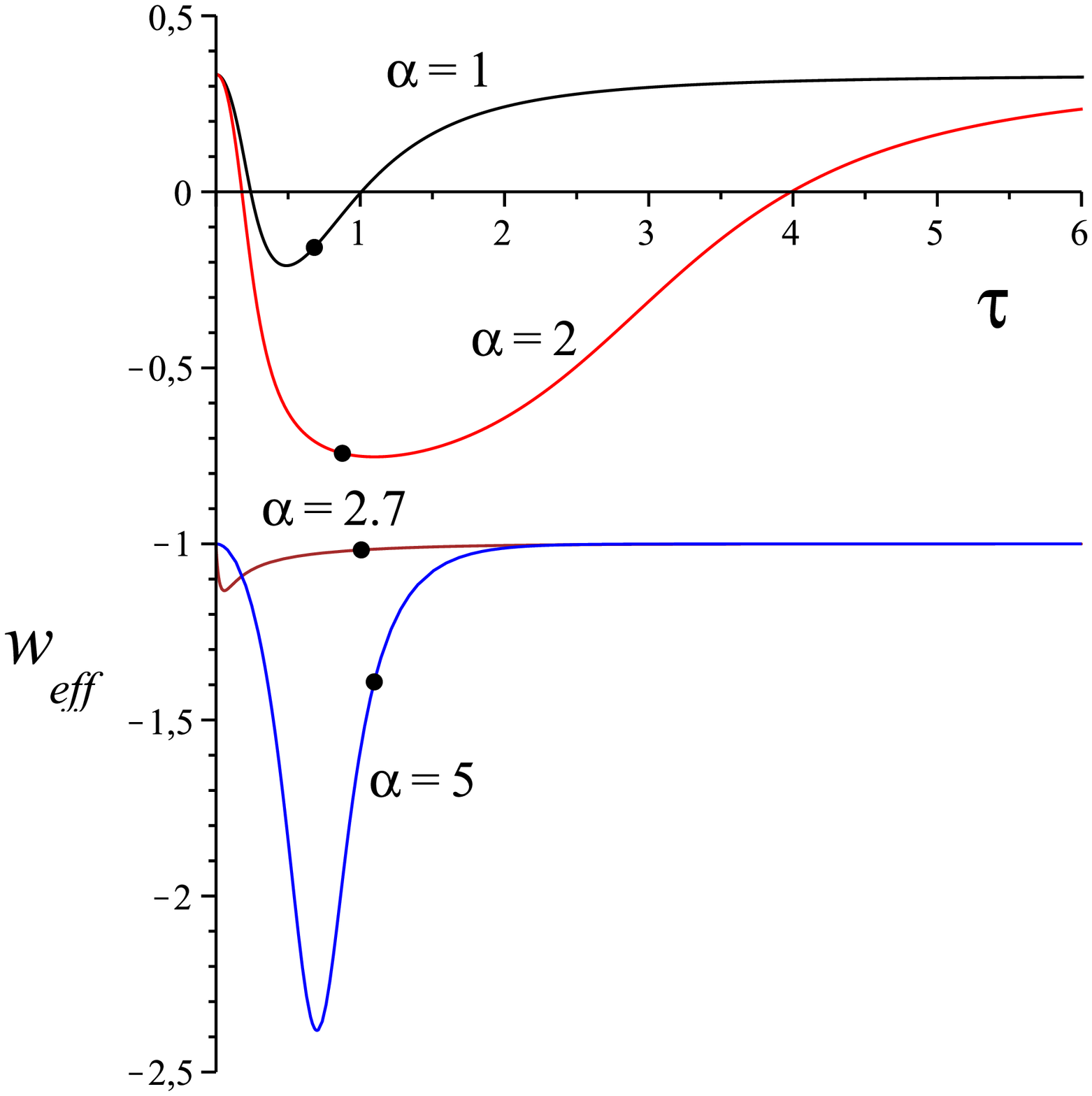}&\qquad
\includegraphics[scale=0.3]{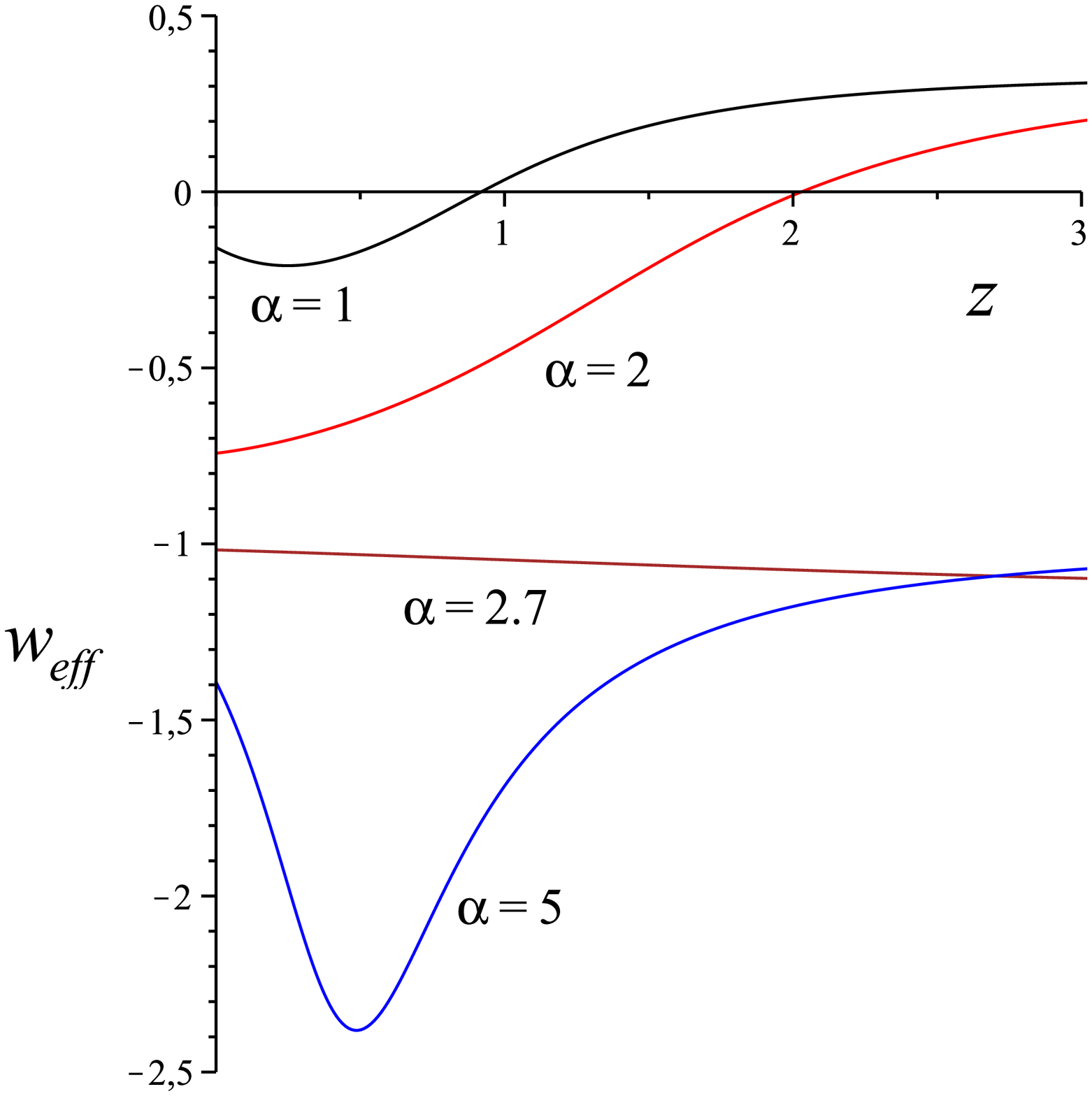}\\[.2cm]
\mbox{(a)} &\qquad \mbox{(b)}\cr
\end{array}
$\\
\end{center}
\caption{The evolution of the effective $w_{\rm eff}\equiv p_{\rm (sc)}/\rho$ is shown in panel (a) for the same choice of parameters and initial conditions as in Fig. \ref{fig:3}.
Panel (b) shows instead its behavior as a function of the redshift.
}
\label{fig:4}
\end{figure*}

Fig. \ref{fig:4} then shows the behavior of the effective $w_{\rm eff}$ both as a function of time and as a function of the redshift, which is related to the scale factor in the standard way, i.e.,
\beq
1+z=\frac{a_0}{a}=\frac{1}{x}\,.
\eeq
The curves for $\alpha=1$ and $\alpha=2$ approach the value $w_{\rm eff}=1/3=w_{\rm(in)}$, whereas those for $\alpha=2.7$ and $\alpha=5$ go to the value $w_{\rm eff}=-1$ corresponding to the equilibrium solutions for the SC density.
In particular, the effective equation of state for $\alpha=2.7$ is $p_{\rm (sc)}\sim-\rho$ at all times, so mimicking quite well the effect of the cosmological constant.
Therefore, we expect the corresponding SC cosmology to be very close to the standard model, as we will show in the next section. 
We have also checked the fulfillment of the energy conditions for the above choice of parameters.
The strong energy condition turns out to be satisfied all along the evolution for small values of $\alpha$ ($\alpha\lesssim1$) only, whereas the null energy condition fails for $\alpha\gtrsim5$.
[Recall that the $\Lambda$CDM model satisfies the null energy condition, but not the strong one.]

Finally, the expression (\ref{qdef}) for the deceleration parameter becomes
\beq
\label{qdefnew}
q=\frac{\Omega_{\rm (sc)}}{2}+\frac32\frac{p_{\rm (sc)}}{\rho_{\rm (crit)}}+\frac{\Omega_m}{2}-\Omega_\Lambda\,,
\eeq
due to the inclusion of the matter density contribution (\ref{matterden}) to the acceleration equation (\ref{accel}).


\section{Observational tests}

\begin{figure}
\begin{center}
\includegraphics[scale=0.4]{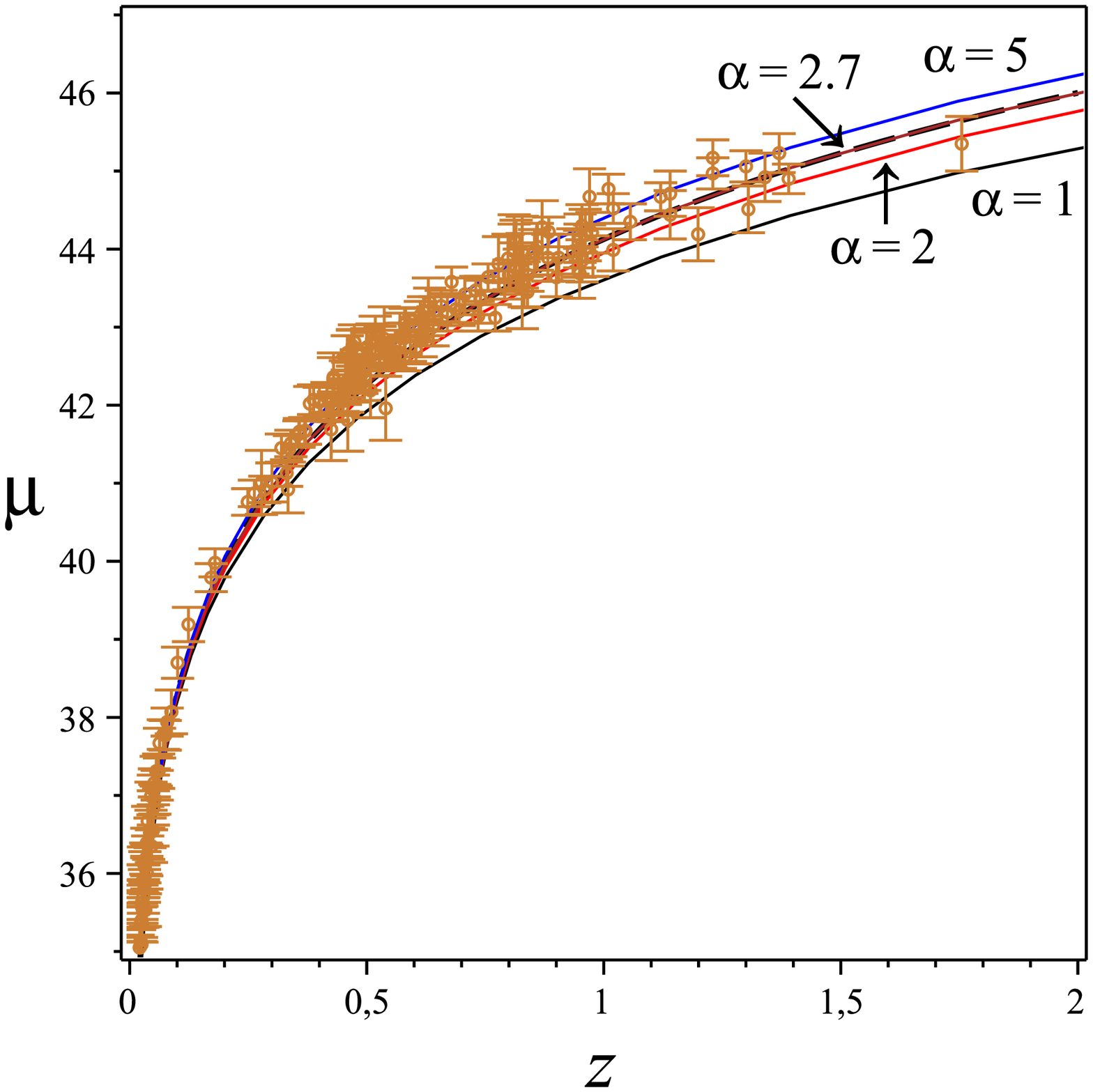}
\end{center}
\caption{Distance modulus vs redshift for different values of the parameter $\alpha=[1,2,2.7,5]$.
Data are taken from Ref. \cite{riess} (gold sample).
The model equations (\ref{eqsvstau1}), (\ref{eqsvstau2new}) and (\ref{eqr}) have been numerically integrated for the same choice of parameters and initial conditions as in Fig. \ref{fig:3}, plus the additional condition $r(\tau_0)=0$ (i.e., $d_c(\tau_0)=0$).
For the Hubble constant we have used the value $H_0=71\,\, {\rm km}\,\,{\rm s}^{-1}\,\,{\rm Mpc}^{-1}$ \cite{komatsu}.
The curve for $\alpha=2.7$ is practically superimposed to the $\Lambda$CDM one (thick black dashed curve).
}
\label{fig:5}
\end{figure}

The distance-redshift relation of SNe Ia is one of the most powerful tools available in observational cosmology.
In Fig. \ref{fig:5} below we compare the fits of the supernova data (gold sample of Ref. \cite{riess}), obtained by plotting the distance modulus $\mu$ versus redshift $z$, both in a SC cosmology without vacuum energy and according to the concordance model.   
The distance modulus is defined by
\beq
\mu=5\log\frac{d_L}{\rm Mpc}+25\,,
\eeq
in terms of the luminosity distance
\beq
d_L=(1+z)\frac{d_H}{\sqrt{|\Omega_{k,0}|}}\Sigma_k\left(\sqrt{|\Omega_{k,0}|}\frac{d_c}{d_H}\right)\,, 
\eeq
where $d_H=1/H_0$ is the Hubble distance and $d_c$ is the comoving distance. 
The comoving distance is defined by $d_c=a_0r$, where $r$ is obtained by integrating the radial null 
geodesic equation $dr/dt=-1/a$ for a light signal emitted at a certain time in the past, by a galaxy 
comoving with the cosmic fluid and received at the present time (i.e., $r=0$ at $t=t_0$).


In the case of the concordance model, the comoving distance is given by (see, e.g., Ref. \cite{peebles})
\begin{eqnarray}
d_c&=&d_H\int_0^z\frac{dz'}{E(z')}\,, \\
E&\equiv&\frac{H}{H_0}=\sqrt{\Omega_{\Lambda,0}+\Omega_{k,0}(1+z)^2+\Omega_{m,0}(1+z)^3}\,. \nonumber
\end{eqnarray}
In the case of a SC cosmology, instead, we have to add the following equation to the system (\ref{eqsvstau1}) and (\ref{eqsvstau2new})
\beq
\label{eqr}
\frac{d}{d\tau}\left(\frac{d_c}{d_H}\right)=-\frac{1}{x}\,.
\eeq
Subsequently, we numerically solve them all together, with the further initial condition $r(\tau_0)=0$ (i.e., $d_c(\tau_0)=0$).
The resulting curve for $\alpha=2.7$ is practically superimposed to the $\Lambda$CDM one.

In order to measure the goodness of fit one can use the method of least squares, which consists in minimizing the function
\beq
\chi^2(\lambda_A)=\sum_{i=1}^n\frac{[\mu(z_i; \lambda_A)-\mu_i]^2}{\sigma_i^2}
\eeq
with respect to the whole set of parameters $\lambda_A=(\alpha,g,w_{\rm (in)},H_0,\Omega_{k,0},\Omega_{m,0})$ of the model.
The $n$ data points $(z_i,\mu_i)$ with errors $\sigma_i$ as inferred from the chosen supernova data set are thus compared with the corresponding expected values of the distance modulus at a given redshift $z=z_i$ for each parameter choice. 
We list in Table \ref{tab:1} the results of the $\chi^2$ statistics for varying $\alpha$ and fixed values of the remaining parameters as in Fig. \ref{fig:3}, showing that the best fit is for $\alpha\simeq4$.
A more accurate analysis would require determining the most likely values as well as the confidence intervals for all parameter sets used in our model by constructing the corresponding likelihood function, but it is beyond the aim of the present work.

\begin{table}
\caption{
The results of the $\chi^2$ analysis applied to the 184 supernova data set from the gold sample \cite{riess} with redshift greater than $z\approx0.023$. The value of  Hubble constant we used is different from the best fitting value of Ref. \cite{riess}, leading to a greater reference value of $\chi^2_{\Lambda{\rm{CDM}}}\approx331$.
}
\begin{ruledtabular}
\begin{tabular}{cccc}
&$\alpha$&$\chi^2/\chi^2_{\Lambda{\rm{CDM}}}$&\\
\hline
& 1 & 3.79 &\\
& 2 & 1.59 &\\
& 2.7 & 0.97 &\\
& 3 & 0.83 &\\
& 4 & 0.66 &\\
& 5 & 0.68 &\\
& 6 & 0.73 &\\
& 7 & 0.79 &\\
& 8 & 0.84 &\\
& 9 & 0.88 &\\
& 10 & 0.91 &\\
\end{tabular}
\end{ruledtabular}
\label{tab:1}
\end{table}

\begin{figure}
\begin{center}
\includegraphics[scale=0.4]{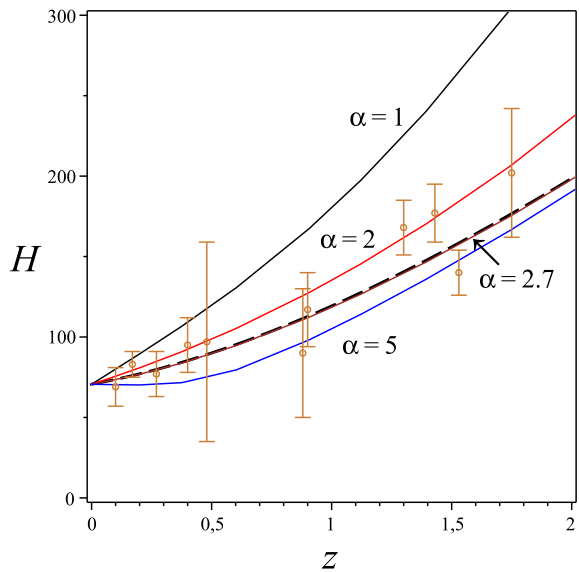}
\end{center}
\caption{Hubble parameter as a function of the redshift for different values of the parameter $\alpha=[1,2,2.7,5]$.
Data are taken from Ref. \cite{simon}.
The prediction from the $\Lambda$CDM model is also shown (thick black dashed curve).
The Hubble parameter is expressed in units of ${\rm km}\,\,{\rm s}^{-1}\,\,{\rm Mpc}^{-1}$ (the value of $H_0$ is the same as in Fig. \ref{fig:5}).
}
\label{fig:6}
\end{figure}

Observational Hubble parameter data have been measured through the aging of passively evolving galaxies \cite{simon} and baryon acoustic oscillations \cite{percival}.
In Fig. \ref{fig:6} we show how the fit of the relation Hubble parameter vs redshift, obtained by using the same set of parameters as in Fig. \ref{fig:3}, is in agreement with the $\Lambda$CDM prediction, despite the small size of the dataset.


\begin{figure}
\begin{center}
\includegraphics[scale=0.4]{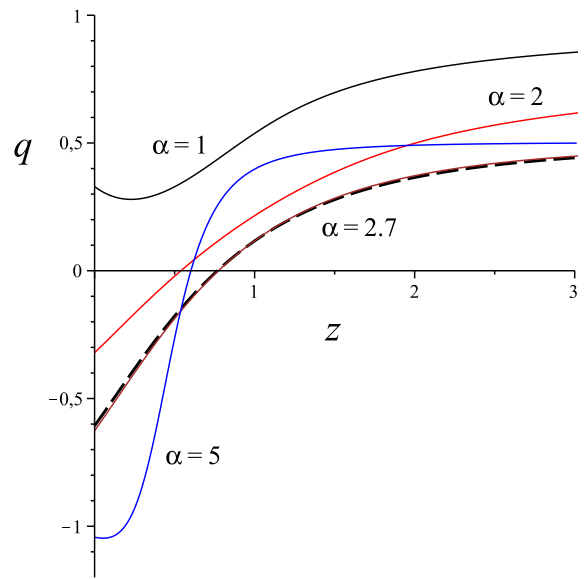}
\end{center}
\caption{Deceleration parameter as a function of the redshift for different values of the parameter $\alpha=[1,2,2.7,5]$.
The choice of parameters as well as initial conditions is the same as in Fig. \ref{fig:3}.
The curve corresponding to the $\Lambda$CDM model is also shown for comparison (thick black dashed curve).
}
\label{fig:7}
\end{figure}

\begin{figure*}
\begin{center}
$\begin{array}{cc}
\includegraphics[scale=0.3]{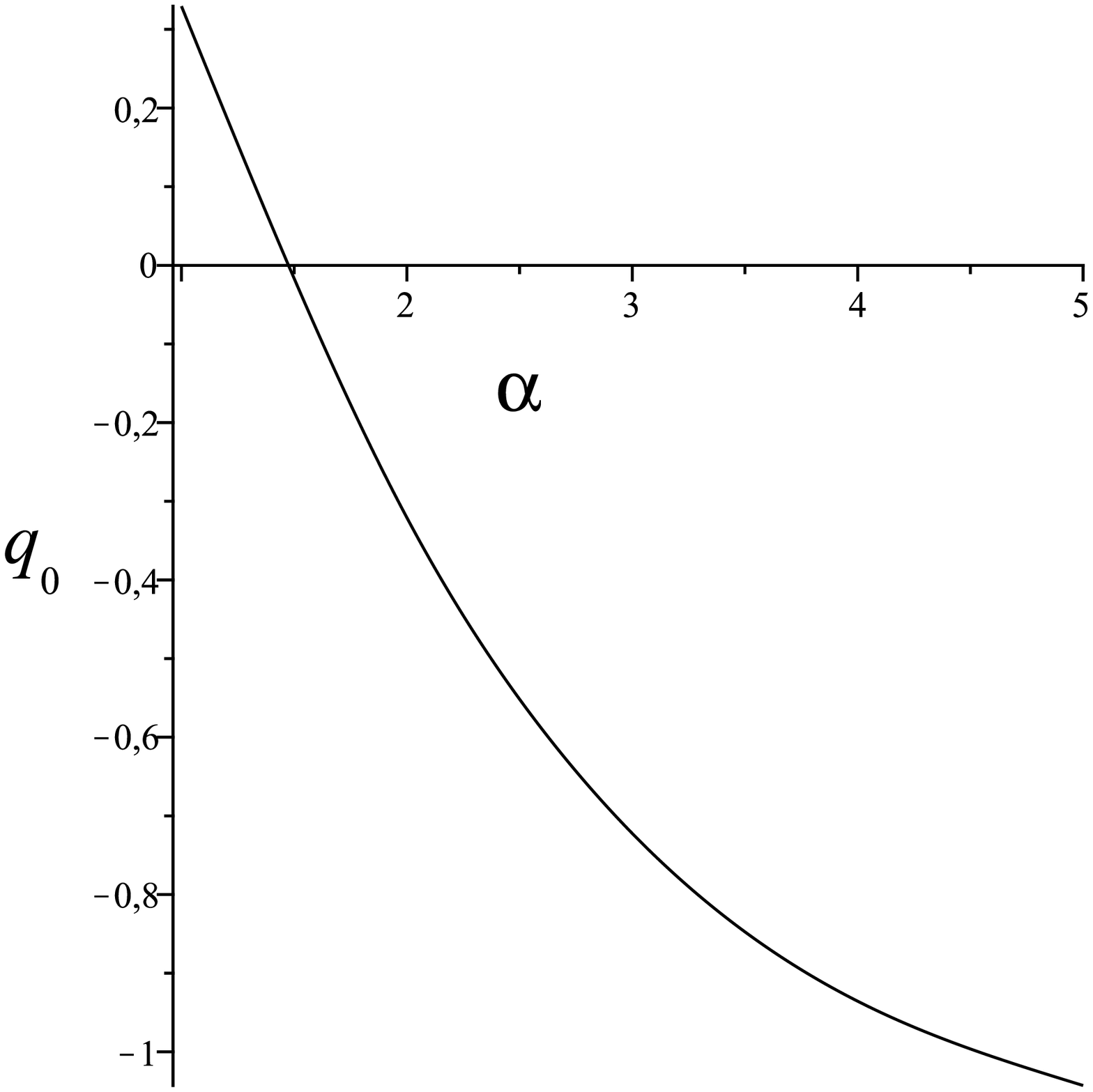}&\qquad
\includegraphics[scale=0.3]{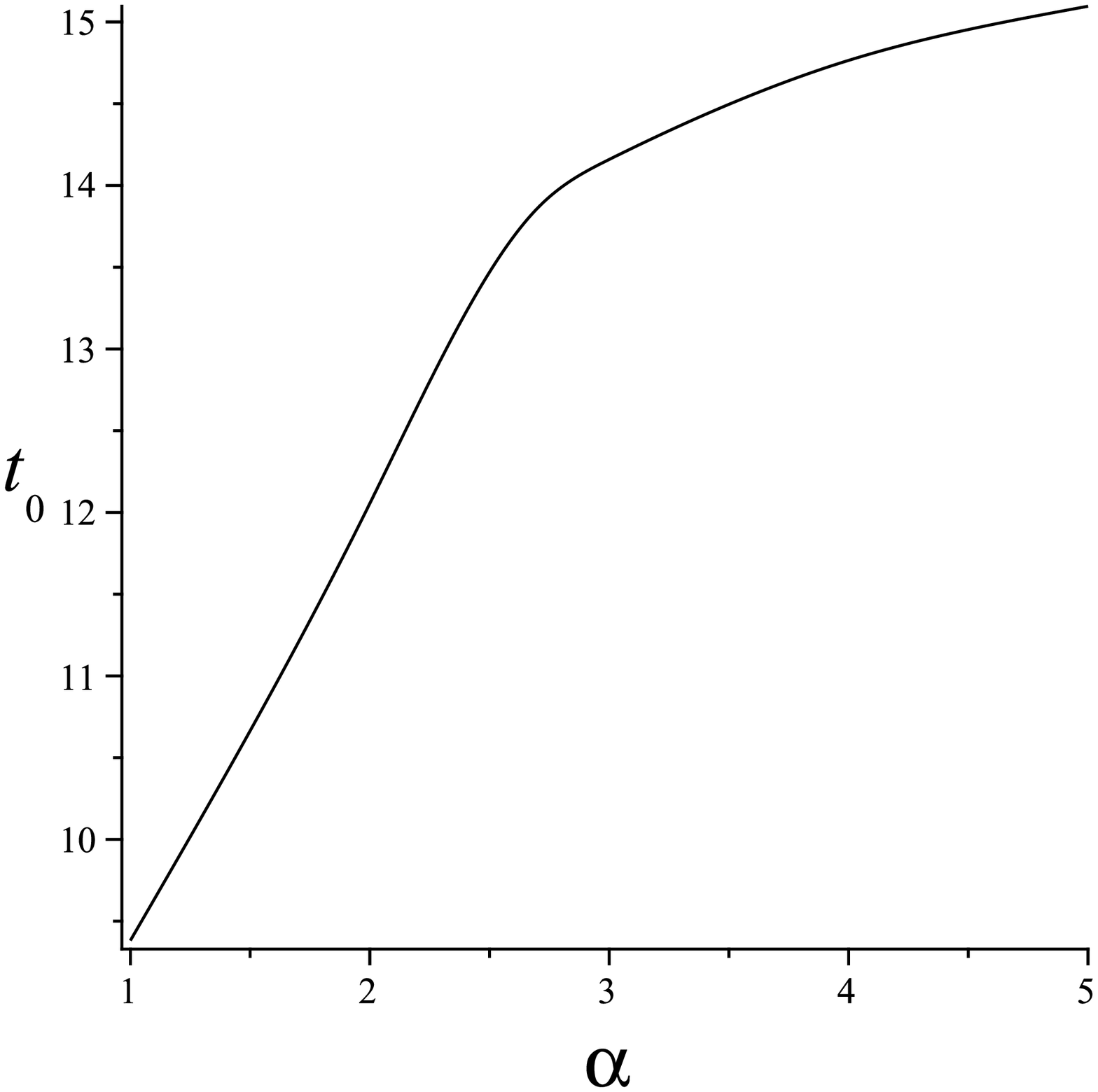}\\[.2cm]
\mbox{(a)} &\qquad \mbox{(b)}\cr
\end{array}
$\\
\end{center}
\caption{Present day value of the deceleration parameter $q_0$ and age of the universe $t_0=\tau_0/H_0$ (expressed in Gyrs) as functions of the parameter $\alpha$.
The choice of parameters as well as initial conditions is the same as in Fig. \ref{fig:3}.
}
\label{fig:8}
\end{figure*}

Fig. \ref{fig:7} shows the behavior of the deceleration parameter (\ref{qdefnew}) as a function of the redshift for different values of the parameter $\alpha$.
The curve corresponding to the $\Lambda$CDM model with  
\beq
q=\frac{1}{H}\frac{dH}{dz}(1+z)-1
\eeq
is also shown for comparison.
Finally, for the same parameter choice as above, one can evaluate the present-day value of the deceleration parameter $q_0$ as well as the age of the universe $t_0=\tau_0/H_0$.
For instance, for $\alpha=2.7$ we obtain $q_0\approx-0.63$ and $t_0\approx13.86$ Gyrs, respectively, in good agreement with current estimates ($q_0=-0.67\pm0.15$ and $t_0=13.75\pm0.17$ Gyrs) \cite{komatsu}, having assumed for the Hubble constant the value $H_0=71\,\, {\rm km}\,\,{\rm s}^{-1}\,\,{\rm Mpc}^{-1}$.
The dependence of both $q_0$ and $t_0$ on the parameter $\alpha$ is also shown in Fig. \ref{fig:8}.


\section{Stability analysis}

During the evolution, the cosmological fluid could suffer the formation of small inhomogeneities due to the development of density gradients as well as the growth of gravitational instabilities which may invalidate the hydrodynamic description. 

At a microscopical level, in the Shan-Chen model phase-separation is triggered by attractive interactions 
between neighboring cells in the lattice.
Attractive interactions enhance density gradients and promote
a subsequent progressive steepening of the interface, eventually
taking the system to a density blow up. In dense fluids and liquids
such density blow up is prevented by hard-core repulsive forces,
which stop the indefinite build-up of density gradients, thereby
stabilizing the fluid interface. As discussed in the Appendix, in the SC model, such a stabilizing effect is obtained by imposing
a saturation of the intermolecular attraction for densities above
a reference value. This is an effective form of ``asymptotic freedom,'' which implies vanishing interactions at short-distance (high-density).

\begin{figure*}
\begin{center}
$\begin{array}{cc}
\includegraphics[scale=0.3]{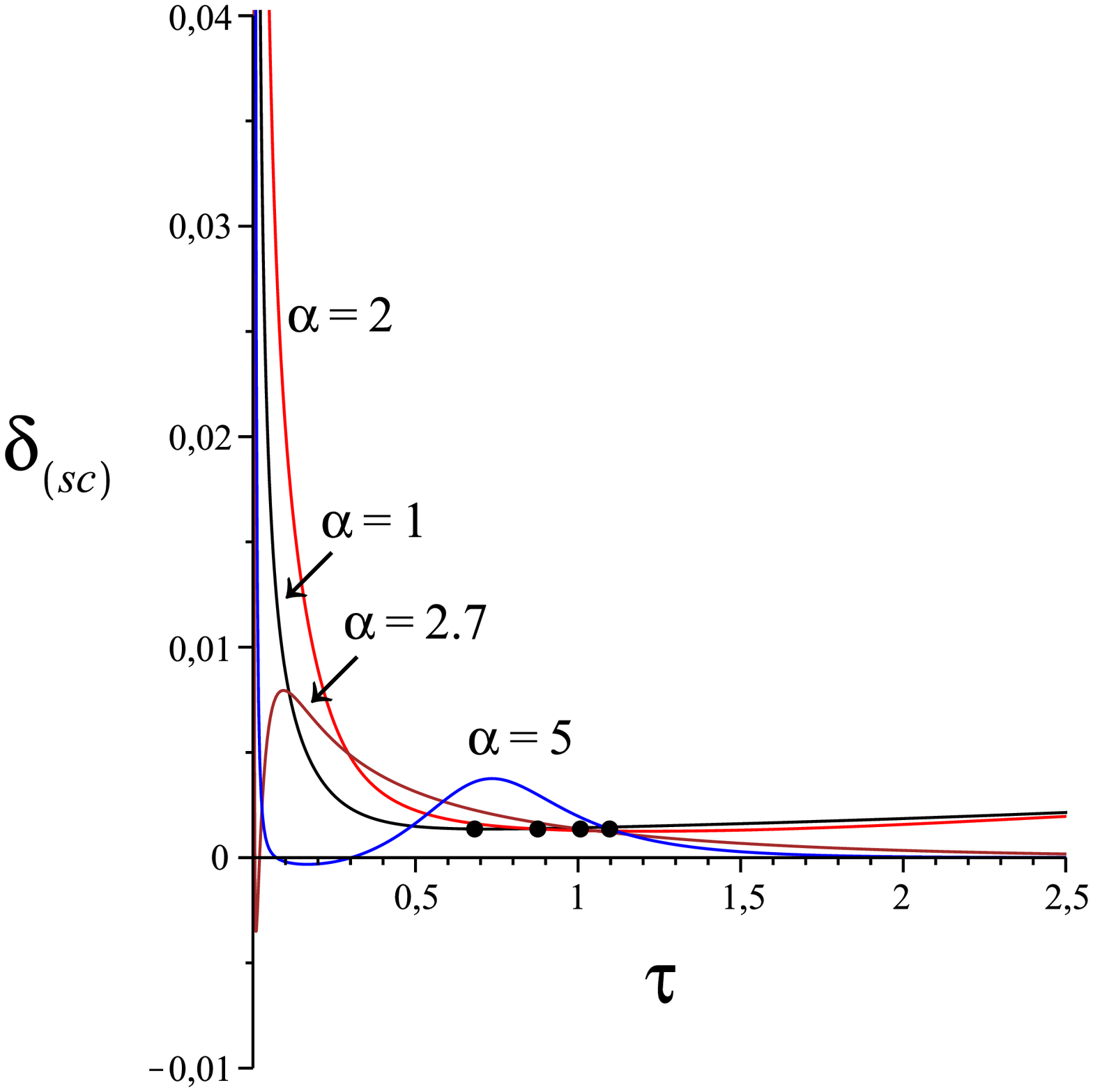}&\qquad
\includegraphics[scale=0.3]{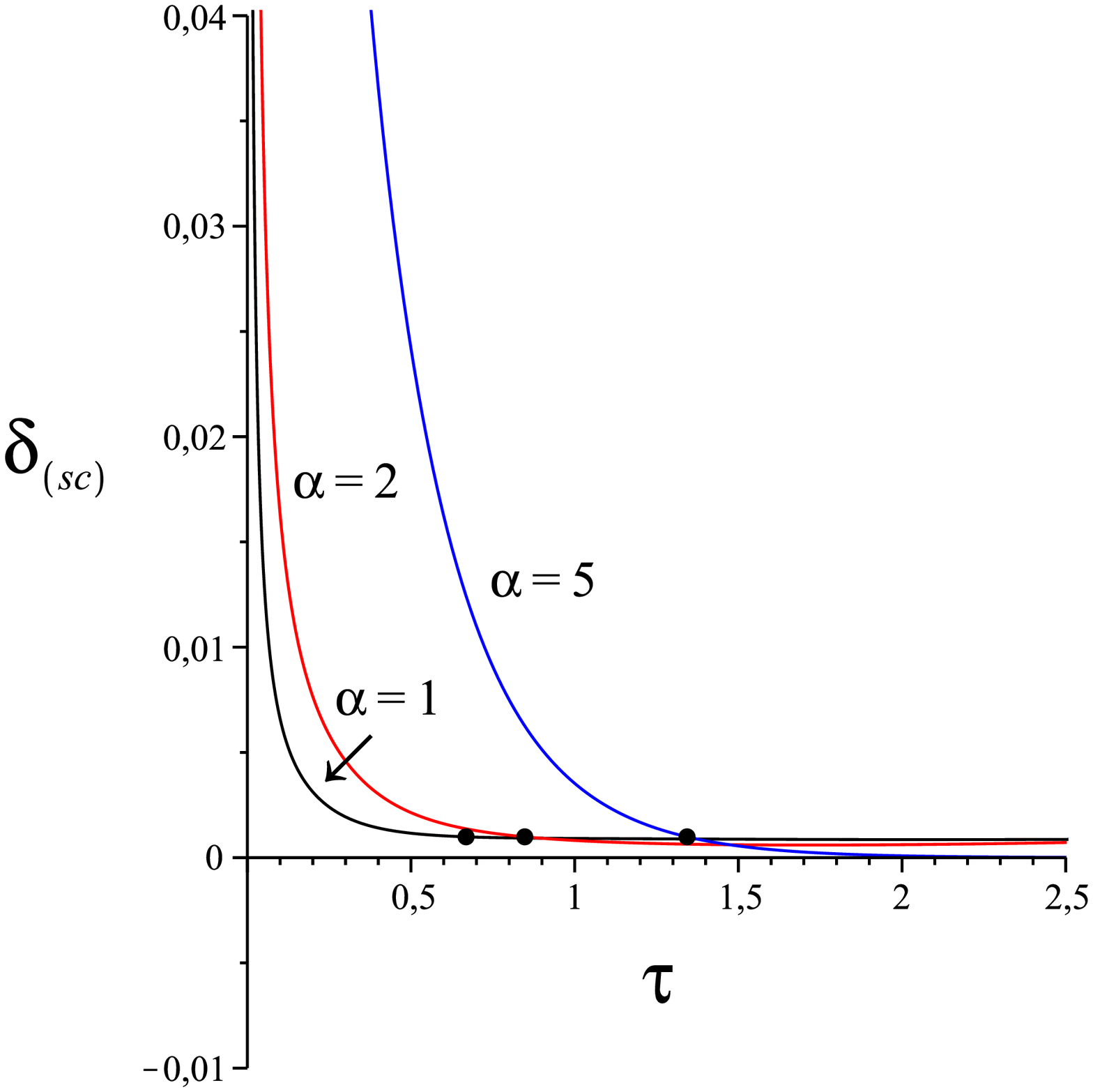}\\[.2cm]
\mbox{(a)} &\qquad \mbox{(b)}\cr
\end{array}
$\\
\end{center}
\caption{The evolution of the density contrast associated with the SC fluid is shown in panel (a) for different values of the parameter $\alpha$.
The linear perturbation equations (\ref{perteqs2}) have been numerically integrated with a representative value of $\tilde k=10^{-2}$ and initial conditions $\delta\xi(\tau_0)=\delta\tilde u(\tau_0)=\tilde\psi(\tau_0)=\delta\tilde\rho_m(\tau_0)=10^{-3}$. 
The choice of remaining parameters as well as initial conditions for the background SC density $\xi$ and scale factor $x$ is the same as in Fig. \ref{fig:3}.
Panel (b) corresponds, instead, to a simple SC model without any additional component in the cosmological fluid.
The choice of parameters as well as initial conditions in this case is the same as in Fig. \ref{fig:2}.
Integrating backward in time shows that now the system becomes soon unstable for every value of $\alpha$.
The presence of a pressureless matter component has then a stabilizing effect on the whole system for $\alpha\gtrsim2$.
In fact, the perturbation undergoes oscillations remaining bounded all the way up to very early times and vanishes at late times. 
}
\label{fig:9}
\end{figure*}

Therefore, a perturbation analysis of our model is in order.
Below, we study scalar linear perturbations of a FRW universe with a SC fluid plus a pressureless matter component in the synchronous gauge, following Ref. \cite{weinberg}.
The set of equations for scalar perturbations is given by
\begin{eqnarray}
\label{perteqs}
0&=&\delta p_{\rm(sc)}+\partial_t[(\rho+p_{\rm(sc)})\delta u]+\frac{3\dot a}{a}(\rho+p_{\rm(sc)})\delta u\,,\nonumber\\
0&=&\delta\dot\rho+\frac{3\dot a}{a}(\delta\rho+\delta p_{\rm(sc)})+\nabla^2\left[a^{-2}(\rho+p_{\rm(sc)})\delta u\right]\nonumber\\
&&+(\rho+p_{\rm(sc)})\psi\,,\nonumber\\
0&=&\dot\psi+\frac{2\dot a}{a}\psi+4\pi(\delta\rho+3\delta p_{\rm(sc)}+\delta\rho_m)\,,\nonumber\\
0&=&\delta\dot\rho_m+\frac{3\dot a}{a}\delta\rho_m+\rho_m\psi\,,
\end{eqnarray}
where $\rho$ and $p_{\rm(sc)}$ are the background values for the SC density and pressure, $\delta\rho$ and $\delta p_{\rm(sc)}$ the corresponding first order perturbations, $\rho_m$ and $\delta\rho_m$ the matter density and its perturbation, $\delta u$ the perturbed scalar velocity potential, and $\psi$ a suitable combination of metric perturbations (not to be confused with the SC extra pressure term in Eq. (\ref{pscdef})).
The (scalar) anisotropic stress tensor for the SC fluid has been assumed to be zero.
Fourier-transforming all perturbation quantities, one obtains the evolution equations for the corresponding amplitudes, with $\nabla^2\to-k^2$ and comoving wave number $k$.
The resulting set of equations in dimensionless form is given by
\begin{eqnarray}
\label{perteqs2}
0&=&w_{\rm (in)}\delta{\mathcal P}_{(\alpha,g)}+(\xi+w_{\rm (in)}{\mathcal P}_{(\alpha,g)})\frac{d(\delta\tilde u)}{d\tau}\nonumber\\
&&+\left[\frac{d\xi}{d\tau}+w_{\rm (in)}\frac{d{\mathcal P}_{(\alpha,g)}}{d\tau}+\frac{3}{x}\frac{dx}{d\tau}(\xi+w_{\rm (in)}{\mathcal P}_{(\alpha,g)})\right]\delta\tilde u\,,\nonumber\\
0&=&\frac{d(\delta\xi)}{d\tau}+\frac{3}{x}\frac{dx}{d\tau}(\delta\xi+w_{\rm (in)}\delta{\mathcal P}_{(\alpha,g)})\nonumber\\
&&+(\xi+w_{\rm (in)}{\mathcal P}_{(\alpha,g)})\left(-\frac{\tilde k^2}{x^2}\delta\tilde u+\tilde\psi\right)\,,\nonumber\\
0&=&\frac{d\tilde\psi}{d\tau}+\frac{2}{x}\frac{dx}{d\tau}\tilde\psi+\frac32(\delta\xi+3w_{\rm (in)}\delta{\mathcal P}_{(\alpha,g)}+\delta\tilde\rho_m)\,,\nonumber\\
0&=&\frac{d(\delta\tilde\rho_m)}{d\tau}+\frac{3}{x}\frac{dx}{d\tau}\delta\tilde\rho_m+\frac{\Omega_{m,0}}{x^3}\tilde\psi\,,
\end{eqnarray}
where 
\beq
\delta{\mathcal P}_{(\alpha,g)}=\frac{\partial{\mathcal P}_{(\alpha,g)}}{\partial\xi}\delta\xi
=\frac{c_s^2}{w_{\rm (in)}}\delta\xi\,,
\eeq
$\delta\tilde u=H_0\delta u$, $\tilde\psi=\psi/H_0$, $\delta\tilde\rho_m=\delta\rho_m/\rho_{\rm (crit),0}$ and $\tilde k=k/(a_0H_0)$.
Numerically integrating the above system of equations together with Eqs. (\ref{eqsvstau1}) and (\ref{eqsvstau2new}) gives the evolution of the density contrast $\delta_{\rm(sc)}\equiv\delta\xi/\xi$ associated with the SC fluid.
We assume here initial conditions such that, at the present time, the values of the perturbation quantities are small enough, i.e., $\delta\xi(\tau_0)\sim\delta\tilde u(\tau_0)\sim\tilde\psi(\tau_0)\sim\delta\tilde\rho_m(\tau_0)\ll1$, in agreement with the observational evidence for a smooth dark energy component.
Integration of the perturbation equations is then performed both backward and forward in time, exactly as is the case of Eqs. (\ref{eqsvstau1}) and (\ref{eqsvstau2new}). 
Fig. \ref{fig:9} (a) shows that the perturbation remains small over a significant time interval. 
It indefinitely grows as the initial singularity is approached, for all fixed values of $\alpha$.
However, in this limit the perturbative analysis is no longer appropriate.
For those solutions evolving between two equilibrium states ($\alpha\gtrsim2$), the perturbation undergoes oscillations remaining bounded all the way down to very early times and vanishes at late times.  
For $\alpha\lesssim2$, instead, the perturbation increases at late times, since the SC background density $\xi\to0$ there.
Furthermore, in this case the exponential growth in the past starts even before, very close to the present time value, indicating a sudden onset of instability.
It is observed that the presence of a pressureless matter component in the cosmological fluid thus leads to a stabilization of the whole system for $\alpha\gtrsim2$.
This is apparent from Fig. \ref{fig:9} (b), which shows the evolution of the SC density contrast for a plain SC model, without any additional component.
Integrating backward in time shows that the system becomes soon unstable for every value of $\alpha$.
Therefore, the inclusion of a matter component in the model plays a role in contrasting formation of instabilities naturally arising in simple SC fluids.
We note that, although pressureless, the matter component also affects the evolution of the background SC density (\ref{eqsvstau1}) via the equation (\ref{eqsvstau2new}) for the evolution of the scale factor. The details of this non trivial stabilization effect will be deferred to a future study.  
Similarly, we leave to a future investigation also the problem of the possible growing of instabilities if the system is assumed to evolve forward in time starting from adiabatic initial conditions at early times (see, e.g., Ref. \cite{weinberg}), consequently implying a different choice of initial conditions for the associated Eqs. (\ref{eqsvstau1}) and (\ref{eqsvstau2new}).


\section{Concluding Remarks}

We have presented a new class of cosmological models consisting of a FRW universe with a fluid source obeying a non-ideal, Shan-Chen-like equation of state. 
The aim of this study was to explain the today dark energy abundance within a different approach with respect to the standard one, which postulates the existence of a mixture of non-interacting perfect fluids as source of a FRW cosmology, including a cosmological constant as responsible for the accelerated expansion of the universe.   
We have shown that, in the case of a simple model without any additional component in the cosmological fluid, starting from an ordinary equation of state at early times (e.g., satisfying the energy condition typical of a radiation-dominated universe), the SC pressure changes its sign at a certain time in the past and remains negative for a large time interval, including the present epoch.
This implies that the equation of state governing the evolution of the present-day universe is typical of dark energy. 
As a result, such a dark energy component develops, with no need of invoking any cosmological constant.
In order to account for the presence of matter density today we have then added to the SC fluid the contribution due to pressureless matter.
The latter is shown to significantly affect the evolution of the SC density, which exhibits a twofold behavior depending on the parameter choice: it either evolves between two equilibrium states or indefinitely grows as the initial singularity is approached and vanishes at late times.
Furthermore, the additional matter component acts so as to contrast the onset of SC instabilities.
In fact, a first order perturbation analysis reveals that a plain SC model is in general unstable against perturbations, whereas the inclusion of a pressureless matter component has a stabilization effect on the SC fluid, at least for those solutions evolving between two equilibrium states. 
We have also provided some observational tests in support to our model.
More precisely, we have drawn the Hubble diagram (distance modulus vs redshift) as well as the expansion history of the universe (Hubble parameter vs redshift), showing that they are consistent with current astronomical data. 
The model opens up several directions for future investigations, for instance a systematic exploration of the remaining parameters of the model, the analysis of different forms $\psi=\psi(\rho)$ of the Shan-Chen excess pressure field and the inclusion of further additional components in the cosmological fluid, as well as a more accurate stability analysis exploring different initial conditions for the perturbation equations.

\appendix

\section{The Shan-Chen model of non-ideal fluids}

It is well known that non-ideal fluid equations of state, say
of van der Waals type, result from underlying atomic potentials
exhibiting short-range (hard-core) repulsion and long-range (soft-core)
attraction. The prototypical example are Lennard-Jones fluids, whose
spherically symmetric potential takes the so-called $6-12$ form
\beq
V(r) = 4 U\left[ \left(\frac{r}{r_0}\right)^{-12} - \left(\frac{r}{r_0}\right)^{-6}\right]\,,
\eeq
where $r_0$ is the typical
equilibrium intermolecular distance, $U$ the typical strength of 
the interaction and $\zeta=r/r_0$ is a natural dimensionless radial variable.
The short-range $-12$ branch leads to very strong repulsive forces on molecules penetrating
the hard-core region $r<r_0$ (actually $r<2^{1/6}r_0\approx 1.12 r_0$), and consequently to impractically 
short time-steps in the numerical integration of the equations of motion
of molecular fluids. 
To circumvent this problem, and with specific reference to {\it lattice} fluids
for which the time-step is fixed by the lattice size --hence cannot be
reduced on demand-- Shan and Chen \cite{shan-chen} proposed a \lq\lq synthetic" repulsion-free potential.
More precisely, repulsion is replaced by a density-dependent attraction, and
the density dependence is tuned in such a way that attraction becomes vanishingly
small beyond a given density threshold, so as to prevent the onset of instabilities 
due to uncontrolled density pile-up.   
Since high-density implies short spatial separation, the Shan-Chen potential implements
a form of effective ``asymptotic freedom,'' meaning by this that molecules below 
a certain separation behave basically like free particles.  

Mathematically, the Shan-Chen interaction leads to the following pair pseudo-potential 
\beq
\label{SCPOT}
V({\mathbf x},{\mathbf x}')=\psi({\mathbf x})G({\mathbf x}-{\mathbf x}')\psi({\mathbf x}')\,, \qquad
{\mathbf x}'={\mathbf x}+{\mathbf e}_a\,,
\eeq
where ${\mathbf e}_a$ denotes a generic spatial direction in the lattice (the explicit dependence on time of the various functions has been 
omitted here to simplify notation). 
For instance, a typical two-dimensional lattice features one rest particle ($|{\mathbf e}_0 |=0$), 
$4$ nearest-neighbors ($|{\mathbf e}_a | =c_L \Delta t$), 
and $4$ next-nearest-neighbors  ($|{\mathbf e}_a| =c_L \Delta t \sqrt 2)$, $c_L=\frac{\Delta x}{\Delta t}$ being the 
lattice \lq\lq light speed." 

In the above, $\psi({\mathbf x}) = \psi[\rho({\mathbf x})]$ is a local functional of the
fluid density and $G({\mathbf x}-{\mathbf x}')$ is the Green function of the interaction.
For the sake of simplicity, Shan and Chen took $G({\mathbf x}-{\mathbf x}')={\mathcal G}<0$ for 
$|{\mathbf e}_a |>c_L \Delta t \sqrt 2$ and zero
elsewhere, so that ${\mathcal G}<0$ codes for attractive interaction.
The associated force per unit volume of the fluid is then 
\begin{equation}
\label{SCFORCE}
{\mathbf F}({\mathbf x})=-\psi({\mathbf x}){\mathcal G}\sum_a\psi({\mathbf x}+{\mathbf e}_a){\mathbf e}_a\,,
\end{equation}
which equals $-\nabla V$ in the limit $\Delta t \to 0$.

Taylor expansion of the above expression gives
\begin{equation}
\label{Force_approx}
{\mathbf F}({\mathbf x}) = -{\mathcal G} \psi({\mathbf x}) \nabla \psi({\mathbf x}) + O(\Delta t^3)\,,
\end{equation}
where we have taken into account that $\sum_a{\mathbf e}_a^i=0$ and 
$\sum_a{\mathbf e}_a^i{\mathbf e}_a^j= (c_L^2 \Delta t^2/3)\delta^{ij}$.
Higher order terms describe physical properties such as surface tension, which play a crucial role in the
dynamics of complex fluids, and are not discussed here.
Confining our attention to the contribution of the above force to the 
equation of state, it is easy to show that such contribution writes as an
excess pressure of the form (in lattice units $\Delta t = \Delta x = c_L =1)$:
\begin{equation}
\label{pstar}
\frac{p}{c_s^2}-\rho = \frac{\mathcal G}2 \psi^2(\rho)\,.
\end{equation}
Note that for attractive interactions, i.e., ${\mathcal G}<0$, this excess pressure is negative.
The functional form $\psi(\rho)$ was chosen in Ref. \cite{shan-chen} in such a
way as to realize a vapor-liquid coexistence curve:
\begin{equation}
\psi(\rho)= \rho_0  \left(1-e^{-\frac{\rho}{\rho_0}}\right)\,,
\end{equation}
where $\rho_0$ is a reference density, above
which \lq\lq asymptotic freedom" sets in.
The definitions of $\psi$ and ${\mathcal G}$ adopted here slightly differ from those used in Section II in order to follow the notation of the original work \cite{shan-chen}.

It is readily checked that, via the equations $\partial p/\partial\rho=0$ and $\partial^2p/\partial\rho^2=0$, the excess pressure (\ref{pstar}) gives rise to
the following set of critical values $\rho_c = \ln 2$, ${\mathcal G}_c = -4$ and $p_c=(\ln2-1/2)/3 \sim 0.063$ at which phase separation starts-off, having set $\rho_0=1$, for simplicity.
In the low density region, $\rho \ll \rho_0$, $\psi \rightarrow \rho$ and
the Shan-Chen equation of state reduces to $p/c_s^2 = \rho + {\mathcal G} \rho^2/2$, $c_s$ being the sound speed of the ideal fluid. 
This is is clearly unstable for ${\mathcal G}<0$, as it yields
$c_s^2 = \partial_{\rho} p <0$ for $\rho > \rho_{\mathcal G} \equiv 1/|{\mathcal G}|$.
This instability is tamed by letting the Shan-Chen force
go to zero for $\rho \gg \rho_0$. 
In the high density limit, the Shan-Chen equation of state reduces to 
$p/c_s^2 = \rho + \frac{\mathcal G}{2} \rho_0^2$.
Consistently with the formal analogy with ``asymptotic freedom,'' this
equation of state bears a close formal resemblance to the bag model of quark matter
\cite{BAG}.
These considerations suggest that the Shan-Chen model might
have a bearing beyond the purpose of a mere technical trick.

In the cosmological context, $\psi({\mathbf x})$ is best 
interpreted as a scalar field, interacting via gauge quanta, whose propagator is given
by $G({\mathbf x}-{\mathbf x}')$ in Eqs. (\ref{SCPOT})--(\ref{SCFORCE}). 
It is worth noting that non-ideal, \lq\lq exotic"
fluids have been proposed before as models of dark energy, one
popular example in point being the (generalized) Chaplygin gas, with equation
of state 
$p=-A/\rho^{\alpha}$,
$A$ being a positive constant and $0 < \alpha \le 1$ \cite{bento}.
A remarkable property of the Chaplygin model is the fact
of supporting negative pressure, jointly with positive sound speed (squared).  
The Chaplygin gas was derived as an approximation to
a fluid dynamic equation of state, most notably as a mathematical
approximation to compute the lifting force on a wing of an airplane \cite{CHAP}.
Lately, it has been capturing increasing interest within the high-energy and cosmological 
communities in view of its large group of symmetry and the fact
that it can be derived from the Nambu-Goto $d$-brane action in $(d+1,1)$ spacetime \cite{GOTO}.
However, to the best of the authors knowledge, no microscopic
basis for the Chaplygin gas model has been provided as yet.

Interestingly, the Shan-Chen equation of state also supports negative 
pressure regimes, jointly with positive $c_s^2$, 
for values of $|{\mathcal G}|$ sufficiently above $|{\mathcal G}_c|=4$. 
Even though any connection of the Shan-Chen model to
string theory remains totally unexplored at the time of this
writing, we note that its equation of state is grounded 
into a sound microscopic basis, namely, according to the expression (\ref{SCFORCE}),
a scalar field interacting through (short-ranged) gauge quanta. 

Based on the above, it appears reasonable to speculate that
the Shan-Chen fluid, by now a very popular model for investigating
a broad variety of complex flows with phase transitions, might have an 
interesting role to play in cosmological fluid dynamics as well.

\begin{acknowledgements}
D. Gregoris is an Erasmus Mundus Joint Doctorate IRAP Ph.D. student and is supported by the Erasmus Mundus Joint Doctorate Program by Grant Number 2011-1640 from the EACEA of the European Commission.
\end{acknowledgements}

\end{document}